\documentstyle[12pt]{article}
\makeatletter
\begin{document}

\topmargin 0pt

\oddsidemargin -3.5mm

\headheight 0pt

\topskip 0mm \addtolength{\baselineskip}{0.20\baselineskip}
\begin{flushright}
  SNUTP-99-008 \\
  SOGANG-HEP $253/99$\\
  hep-th/9903xxx
\end{flushright}
\vspace{0.5cm}
\begin{center}
  {\large \bf New Gauge Invariant Formulation of the Chern-Simons
Gauge Theory: Classical and Quantal Analysis }
\end{center}
\vspace{0.5cm}

\begin{center}
  Mu-In Park$^{*}$\footnote{Electronic address:
    mipark@physics3.sogang.ac.kr} \\ 
  {\it Center for Theoretical Physics, Seoul National University, \\
    Seoul, 151-742, Korea } \\ 
and \\
Young-Jai Park$^{*}$\footnote{Electronic address: yjpark@ccs.sogang.ac.kr}\\
   $^{*}$ {\it  Department of Physics, Sogang University, \\
    C.P.O. Box 1142, Seoul 100-611, Korea} \\
\end{center}
\vspace{0.5cm}
\begin{center}
  {\bf ABSTRACT}
\end{center}
Recently proposed new gauge invariant formulation of the Chern-Simons
gauge theory is considered in detail. This
formulation is consistent with the gauge fixed formulation.
Furthermore it is found that the canonical (Noether) Poincar\'e
generators are 
not gauge invariant even on the constraints surface and do not satisfy
the Poincar\'e algebra contrast to usual case. It is the improved generators,
constructed from the symmetric energy-momentum tensor, which are
(manifestly) gauge invariant and obey the quantum as well as classical
Poincar\'e algebra. The physical states are constructed
and it is found in the Schr\"odinger picture that unusual gauge {\it
invariant} longitudinal mode of the gauge field is
crucial for constructing the physical wavefunctional which is
genuine to (pure) Chern-Simons theory. In matching to the gauge
fixed formulation, we consider three typical gauges, Coulomb, axial
and Weyl gauges as explicit examples. Furthermore, recent several
confusions about the 
effect of Dirac's dressing function and the gauge fixings are
clarified. The analysis according to old gauge independent formulation
$ a'~ la$ Dirac is summarized in an appendix. 
\vspace{0.5cm} 
\begin{flushleft}
  PACS Nos: 11.10.Ef, 11.10.Lm, 11.15.-q, 11.30.-j \\
March 1999 \\
\end{flushleft}

\newpage

\begin{center}
{ \bf I. Introduction}
\end{center}

In general, there are two approaches in
quantum field theory: the gauge invariant formulation (GIF) and the
gauge fixed formulation (GFF). The latter is the conventional one,
where one chooses a gauge. In the former what we are interested in
this paper on the other hand, one does
not fix the gauge but works with gauge invariant quantities. There are
several methods which achieve this former approach depending on what
one chooses the gauge invariant variables
\cite{Gol:78,Haa:95,Kar:96,Bel:96} and his interests. In this paper,
we consider the formalism $a'~la$ Dirac \cite{Dir:55} which provides
one of the simplest formulation and whose validity is independent on
the space-time dimensions and the treating models in principle. 

The idea of the Dirac's gauge invariant formalism is to describe all
physical effect by the 
manifestly gauge invariant variables, called {\it physical variables}
by introducing Dirac's dressing function $c_k ({\bf x},{\bf y})$: The
physical variables are
defined as the quantities which commute with all first-class 
constraints within the Hamiltonian formulation, which are
believed to be directly measurable ones, such that the gauge fixing
condition needs not be introduced . If this procedure is succeeded,
the model is said to be gauge invariant within that formulation;
usually this was trivial thing at classical level but non-trivial at
quantum level even for the gauge theories which have the gauge invariance
of Lagrangian (density) \cite{Dir:55}.
 
Recently this gauge invariant method using the Dirac's gauge invariant
variables has been of considerable interest in the pertubative analysis
of QED and QCD, especially in relation to the infrared divergence and quark
confinement problems \cite{Lav:93}. However
as for the formalism itself, it is not clear how the results in GIF
can be matched to GFF even though this matching has been considered in
several recent analyzes \cite{Gae:97}. 

Furthermore, a similar gauge independent Hamiltonian analysis
\cite{Ban:92} $\acute{ a}~ la $ Dirac \cite{Dir:55,Dir:64} 
has been recently considered for the
Chern-Simons (CS) gauge theory with matter fields
\cite{Hag:84}. Actually after the CS gauge theory was invented, there
arose several debates about the gauge dependence of the spin and
statistics transmutation 
phenomena for the charged matter fields, since the analysis was
carried out with specific gauge fixing
\cite{Hag:84,Hag:85,Shin:92}. So, with the formulation 
without gauge fixing, one can expect to resolve this debate since one
is not confined to a specific gauge. But the result of the recent
gauge independent analysis for this problem in Ref. \cite{Ban:92} 
was questionable since there was no room for spin transmutation. This was in
sharp contrast to the well-known spin transmutation of GFF
\cite{Hag:84,Hag:85,Shin:92}.

This paper is devoted to a detailed study of our new gauge invariant
Hamiltonian formulation that has been recently suggested as a
resolution of these problems \cite{Park:98a}. In Sec. II, we introduce a
physically plausible assumption for the Poincar\'e transformation of
the Dirac's gauge invariant fields that these fields transform
conventionally to the $space$ and $time$ transformation. As a result
we find a new set of equations for Dirac
dressing function $c_{k}({\bf x},{\bf y})$. However, these fields do not
transform  conventionally under the spatial rotation and Lorentz
boost. In Sec. III, we consider quantization. It is found that the gauge
invariant field operators satisfy the graded commutation (exchange) relations
depending on the dressing, and physical states are constructed as the
products of the gauge invariant field operators with the 
gauge invariant vacuum state algebraically. This is compared to the
physical wavefunctional in the Schr\"odinger picture and it is found,
as a genuine effect of the (pure) CS theory, that the gauge
{\it invariant}
but longitudinal mode of the gauge field $A_i$ is important as well as
the usual gauge $varying$ longitudinal mode which carries full gauge
transformation 
property of $A_i$. Moreover, it is shown that the improved generators,
constructed from the symmetric (Belinfante) energy-momentum tensor
\cite{Bel:40}, which are (manifestly) gauge invariant, obey the
Poincar\'e algebra. But we show that canonical (Noether) Poincar\'e
generators are not gauge invariant ``even on the constraints surface''
and do not satisfy the Poincar\'e algebra. These results are valid
even at the quantum level as well as as the classical ones. The
inequivalence of the improved and canonical generators is
essentially due to the CS term, and is important for genuine spin
transmutation in the relativistic CS gauge theory. Furthermore the
fact that only the symmetric energy-momentum tensor, not the canonical
one, is meaningful is consistent with Einstein's theory of gravity.
In Sec. IV, we provide and explain our recently proposed method which
matches GIF to GFF consistently. The Coulomb, axial and Weyl gauges
are considered as explicit examples. Moreover we clarify several
confusions which result from the misunderstanding the gauge fixing
kernel and the dressing function. 
Sec. V is devoted to the discussion and summary. As discussion, we
have considered the manifestly gauge invariant action and its
possible generalization which is connected to the known 
equivalence of the self-dual and Maxwell-CS theories. In Appendix {\bf A}, we
consider the old Dirac formalism for the determination of the
extended Poincar\'e generators and we find that
this formulation is valid only when one neglects the singular boundary
terms. In Appendix {\bf B}, we derive the physical wavefunction of the
Maxwell-CS theory in the Schr\"odinger picture in our context and with
emphasis on the difference to the (pure) CS theory.
In Appendix {\bf C}, we present the proof of the master formula
for the matching of GIF and GFF.

\begin{center}
{ \bf II. New gauge invariant formulation} \\
{\bf A. Dirac's gauge invariant variables }
\end{center} 

Our model is the Abelian CS gauge theory with massive relativistic
complex scalars \cite{Ban:92,Hag:84} which is described by the
Lagrangian density 
\begin{eqnarray}
\label{eq:CS_action}
{\cal L}= \frac{\kappa}{2} \epsilon^{\mu \nu
  \rho}A_{\mu}\partial_{\nu}A_{\rho} 
+(D_{\mu}\phi)^{*}(D^{\mu}\phi)-m^{2} \phi^{*} \phi ,
\end{eqnarray}
where $\epsilon^{012}=1$, $g_{\mu \nu}$=diag(1,--1,--1), and $D_{\mu}=
\partial_{\mu}+iA_{\mu}$. This Lagrangian density is invariant up to
the total divergence under the gauge transformations
$
\phi (x) \rightarrow exp[-i\Lambda (x)] \phi (x),
A_{\mu} (x) \rightarrow A_{\mu} (x) +\partial_{\mu}\Lambda (x), 
$
where $\Lambda$ is a well-behaved function such that $\epsilon^{\mu
  \nu \lambda} \partial_{\mu} \partial_{\nu} \Lambda=0$.
As a reflection of this symmetry, there are
the first-class constraints
\begin{eqnarray}
T_0 &\equiv & \pi_0 \approx 0, \\
\label{eq:T}
T &\equiv & J_0-\kappa B \approx 0 , 
\end{eqnarray}
which are the primary and secondary constraints, respectively, and
there is also a second class-class constraints 
$
 T_i \equiv \pi_i -\frac{\kappa}{2} \epsilon_{ij}A^{j}\approx 0 ~(i=1,2),
$
which results from the symplectic structure of (\ref{eq:CS_action}) in
the Dirac's canonical formalism. Here, $J_0$ is the time component of
the conserved matter current 
$
J_{\mu}=i [ (D_{\mu} \phi)^{*} \phi -\phi^{*} D_{\mu}
\phi]
$
and $B= \epsilon_{ij}\partial_{i} A^{j}$ is the magnetic field.
But all of them are not crucial in the development of our formulation:
It is found that only the secondary constraints $T \approx 0$ is
the non-trivial one. Actually, the Faddeev-Jackiw (FJ) (or
symplectic) bracket method does the work properly and in this method the
basic (equal time) Poisson brackets (called FJ or
  symplectic) brackets \cite{Fad:88}) become 
\begin{eqnarray}
\label{eq:basic_bracket}
&&\{ A^i({\bf x}), A^j({\bf y}) \} =\frac{1}{\kappa} \epsilon^{ij}
 \delta^{2} ({\bf x}-{\bf y}), \nonumber \\
&&\{ \phi({\bf x}), \pi ({\bf y}) \}=
 \{\phi^{*} ({\bf x}), \pi^{*}({\bf y}) \} = 
\delta^{2}({\bf x}-{\bf y}), \\
&&\mbox{others~vanish} \nonumber
\end{eqnarray}
with $\pi=(D_{0}\phi)^{*}$, $\pi^{*} =D_{0} \phi$, and there remains
only the (Gauss' law) constraint $T ({\bf x})\approx 0 $: In this FJ
method, the primary constraints of the Dirac bracket 
method, $T_{0} \approx 0,~T_i \approx 0$, need not be introduced.

Now in order to develop the manifestly gauge invariant Hamiltonian
formulation we introduce the following variables
\begin{eqnarray}
\label{eq:phys_variable}
&&\hat{\phi}({\bf x}) \equiv \phi({\bf x}) exp\left(iW({\bf
    x})\right),\nonumber \\ 
&&\hat{\pi}({\bf x}) \equiv \pi({\bf x}) exp\left(-iW({\bf
    x})\right),\nonumber \\ 
&&{\cal A}_{\mu}({\bf x})\equiv{A}_{\mu}({\bf x})-\partial_{\mu}W ({\bf x}),
\end{eqnarray}
and their complex conjugates with 
\begin{eqnarray}
\label{eq:W}
W({\bf x})=\int d^2 {\bf z}~ c_{k}({\bf x}, {\bf z})
A^{k}({\bf z}) . 
\end{eqnarray}
These variables are manifestly gauge invariant, i.e., 
$
\{ T({\bf x}), {\cal F}_{\alpha} ({\bf y}) \} =0,~ {\cal F}_{\alpha}=
({\cal A}_i, \hat{\phi}, \hat{\pi})
$ 
in the Hamiltonian formulation\footnote{In order to include ${\cal
  A}_0$ also in this category, one could introduce $\int d^2 {\bf x}~
  (\partial _0 
  \Lambda )\pi_0 $ as a temporal-gauge transformation with $\{ A_0
  ({\bf x}), \pi_0 ({\bf y})\}=\delta ^2 ({\bf x} -{\bf y})$ in addition
  to (\ref{eq:basic_bracket}).} if the Dirac dressing function $c_{k}({\bf
  x},{\bf z})$ satisfies
\begin{eqnarray}
\label{eq:dressing}
\partial^{k}_{z} c_{k}({\bf x},{\bf z})=-\delta^2({\bf x}-{\bf z}). 
\end{eqnarray}
Here, we note that there are infinitely many solutions of $c_k({\bf
  x},{\bf z})$ which 
satisfy (\ref{eq:dressing}) and the gauge invariance of the variables in
  (\ref{eq:phys_variable}) should be 
understood on each solution hyper-surface but not on the entire
solution space: This will be discussed in detail in Sec. IV.

Moreover, we note that the decomposition of the base gauge fields as
\begin{eqnarray}
\label{eq:new_decomp}
  { A}_i ({\bf x})= {{\cal A}}_i ({\bf x}) + \partial_i W({\bf x})
\end{eqnarray}
is not always the same as the usual decomposition to the transverse and
longitudinal components 
\begin{eqnarray}
\label{eq:tl_decomp}
  {A}_i ({\bf x})= {A}_i^{T} ({\bf x}) +{ A}_i^{L} ({\bf x})
\end{eqnarray}
with $\nabla \cdot  {\bf A}^{T}=0 $ and $\nabla \times  {\bf
  A}^{L}=0 $: Although $({{\cal A}}_i,\partial_i  W )$ is similar to 
$({A}_i^T,{A}_i^L )$ in that the first parts do not gauge
  transform and only the second parts gauge transform, $
  {\cal A}_i$ does not always satisfy the divergence-free condition:
  \begin{eqnarray}
\label{eq:div_AA}
    \partial^i { {\cal A}}_i ({\bf x}) &=& \partial_i {A}_i ({\bf x})+
    \int d^2 {\bf z} 
    ~\nabla ^2 _x c_k ({\bf x}- {\bf z}) A^k ({\bf z})  \\
                     &\neq & 0 .\nonumber 
  \end{eqnarray}
[The transformation between these two decomposition method
will be discussed in detail in Sec. III. {\bf B}; it will be shown also that
${{\cal A}}_i$ satisfies more generalized condition, and this
generalized condition reduces to the divergence-free condition 
for a particular form of $ c_k ({\bf x}- {\bf z})$ in Sec. IV.] On the
other hand, in the usual decomposition (\ref{eq:tl_decomp}), the gauge
invariant variables (\ref{eq:phys_variable}) for the matter fields
are found to be 
\begin{eqnarray}
  \label{eq:phys_variable_usual}
  \hat{\phi}({\bf x}) &=& {\phi}({\bf x}) exp \left[-i \partial^{-1}_i
{A}_i^L ({\bf x}) \right], \nonumber \\
  \hat{\pi}({\bf x}) &=&{\pi}({\bf x})exp \left[i \partial^{-1}_i 
  {A}_i^L ({\bf x}) \right]
\end{eqnarray}
when the zero-mode of the operator $\partial_i \partial_i=\nabla^2$ is
unimportant such that 
$\nabla^{-2} \nabla^2 =1$ [$\partial^{-1}_i \equiv \partial_i
\nabla^{-2}$ and $\nabla^{-2}$ is defined as the ordered equation
$\nabla^2 \nabla^{-2} =1$]\footnote{Recently the replacement $
  {\phi}({\bf x}) \rightarrow {\phi}({\bf x}) e^{i \partial^{-1}_i
    {\bf A}_i^L ({\bf x}) } $, which is the same as what 
  one uses when he wants to remove the gauge dependent 
${\bf A}_i^L$ part in the first order form of the Lagrangian,
has been understood as a Darboux transformation in the context of the
Hamiltonian reduction \cite{Fad:88}
: Although they didn't use $\hat{\phi}$
explicitly, their renamed field $\phi$ in the right-hand side is
nothing but $\hat{\phi}$ in our formulation.} in the Coulomb gauge.
\begin{center}
  {\bf B. Poincar\'e Transformation of gauge invariant fields }
\end{center}

In order to give some physical meaning to the gauge invariant fields
(\ref{eq:phys_variable}), the transformation properties under the
Poincar\'e generators 
should be defined. To this end, let us
consider the (manifestly) gauge invariant Poincar\'e generators which
are expressed only by the gauge invariant fields and ${\cal D}_i
\equiv {\partial}_{i} +i {\cal A}_i$:
\begin{eqnarray}
\label{eq:Poincare_gen}
&&{P}^{0}_{s} = \int d^{2}{\bf x} \left[ |{\hat{\pi}}|^{2}
      +|{\cal D}^{i} {\hat{\phi} } |^{2} +m^{2} |{\hat{\phi}}|^{2}
       \right], \nonumber \\
&&{P}^{i}_{s} = \int d^{2}{\bf x} \left[{\hat{ \pi}}{\cal D}^{i}{\hat{\phi}} 
      +({\cal D}^{i}{\hat{\phi}})^{*} {\hat{\pi}^{*}}
       \right], \nonumber \\
&&{M}^{12}_{s} = \int d^{2}{\bf x}~ \epsilon_{ij}x^{i} \left[
  {\hat{ \pi}}{\cal D}^{j}{\hat{\phi}} 
  +({\cal D}^{j} {\hat{\phi}})^{*}  {\hat{\pi}^{*}}
   \right], \nonumber \\
&&{M}^{0i}_{s} = x^{0} {P}^{i}_{s}-
 \int d^{2}{\bf x}~ x^{i}\left[|{\hat{ \pi}}|^{2} 
 +|{\cal D}^{j}{\hat{\phi}}|^{2} +m^{2} |{\hat{\phi}}|^2 \right].
\end{eqnarray}
These are $improved$ generators following the terminology of Callan
$et~ al.$ \cite{Cal:70} constructed from the symmetric (Belinfante)
energy-momentum tensor $T^{\mu \nu}_{s}$ \cite{Bel:40}
:
\begin{eqnarray*}
T^{\mu \nu}_{s}({\bf x}) &\equiv& \left. \frac{\delta I}{\sqrt{g}
\delta g_{\mu \nu} ({\bf x})} \right|_{g _{\mu \nu}
\rightarrow \eta_{\mu \nu}} \nonumber \\
&=&(D^{\mu} \phi)^{*}(D^{\nu} \phi) + (D^{\nu} \phi)
(D^{\mu} \phi)^{*} -\eta^{\mu \nu}\left[(D^{\rho} \phi)^{*}
  (D_{\rho} \phi)-m^{2} \phi^{*} \phi \right], \nonumber \\
&=&({\cal D}^{\mu} \hat{\phi})^{*}({\cal D}^{\nu} \hat{\phi}) + ({\cal
  D}^{\nu} \hat{\phi})
({\cal D}^{\mu} \hat{\phi})^{*} -\eta^{\mu \nu}\left[({\cal D}^{\rho}
  \hat{\phi})^{*} 
  ({\cal D}_{\rho} \hat{\phi})-m^{2} \hat{\phi}^{*} \hat{\phi} \right], \\
P^{\mu}_{s} &=& \int d^{2}{\bf x}~ T^{0 \mu}_{s}, \\
M^{\mu \nu}_{s} &=& \int d^{2} {\bf x} ~ \left[x^{\mu}T^{0\nu}_{s}
  -x^{\nu}T^{0\mu}_{s} \right],
\end{eqnarray*}
where $\hat{\pi}^*={\cal D}^0 \hat{\phi} =(\partial^0 +i {\cal A}_0 )
\hat{\phi}$ and $I=\int d^3 x {\cal L}$. Though the gauge invariance
of $P^{\mu}_s, M^{\mu \nu}_s$ 
of (\ref{eq:Poincare_gen})
is manifest (at least classically), there can be added 
additional term $\Gamma
\equiv \int
d^2 {\bf x} ~[ u({\bf x}) T({\bf x}) + u_0 ({\bf x}) T_0 ({\bf x})]$,
which is proportional to the (first-class) constraints, to
generate the correct Poincar\'e transformations for the undressed base
fields $\phi$ and $A^{\mu}$ \cite{Dir:64}. However, we note that as far as we
are interested in 
the dynamics of the physically relevant variables of
(\ref{eq:phys_variable}), we do not need this 
additional term $\Gamma$ in the Poincar\'e
generators of (\ref{eq:Poincare_gen}) \cite{Ban:92,Dir:64} 
since $\Gamma$ has the vanishing Poisson
bracket with the gauge invariant variables ${\cal F}_{\alpha}$.
(See Appendix {\bf A} for the fixing the constraints term
$\Gamma$ from the
transformation properties of the undressed base fields {\it a' la}
Dirac.) Then, the generators in (\ref{eq:Poincare_gen}) should
generate the correct transformation for the 
${\cal F}_{\alpha}$; but this will depend on how the variables ${\cal
  F}_{\alpha}$ are defined. 

In the followings, we will define the
gauge invariant {\it canonical} fields ${\cal
  F}_{\alpha}=({\cal A}_i, \hat{\phi}, \hat{\phi}^* )$, which appear
in the Poincar\'e generators, as what
involving  a particle-like object and investigate what
information can be obtained from this definition. 

\begin{center}
{\bf 1. Space and time translations }
\end{center}

First of all, we consider the spatial translation generated by
\begin{eqnarray*}
\{ {\hat{\phi}}({\bf x}),{P}^{j}_{s}\}
&=&\partial ^j \phi ({\bf x}) e^{i W ({\bf x})} + i \phi ({\bf x})  e^{i W
  ({\bf x})}\int d^2 {\bf z} ~ c_k ({\bf x}, {\bf z}) \partial^i _z
A^k ({\bf z}) \nonumber \\ 
&=& \partial^j {\hat{\phi}}({\bf x}) -i {\hat{\phi}}({\bf x})
\int d^2 {\bf z} \left[\partial^j_z c_k ({\bf x},{\bf z}) + \partial^j_x c_k
({\bf x},{\bf z})\right] {A}^k ({\bf z}).
\end{eqnarray*}
From the second to third lines, the integration by parts has been performed
and we have dropped the terms 
$
\int d^2 {\bf z} ~\partial^i_z \left(c_k({\bf
  x},{\bf z}) {A}^k({\bf z})\right)~ \mbox{and}  \int d^2 {\bf z} ~
  \partial^k_z 
\left(c_k({\bf x},{\bf z}) {A}^j({\bf z})\right), 
$
 which vanish for
sufficiently rapidly decreasing integrand.  This seems to show the
translational anomaly due to the second term. ( Following the
terminology of Hagen $et~al.$ 
\cite{Hag:84,Hag:85,Shin:92}, $``$anomaly" means an unconventional
contribution whose origin 
is at the classical level.) However,
this anomaly should not appear in order that
${\hat{\phi}}$ responds conventionally to translations, i.e., $\{
\hat{\phi} ({\bf x}), P^j _s \} =\partial ^j \hat{\phi} ({\bf x})$
like as the 
particle-like object which is described by the local fields: The
usual local fields have no translational anomaly, even though they have
additional terms in the rotation because of their spin or other
  properties.\footnote{Similar assumption has been recently
  considered independently in a different context by Bagan {\it et
    al.} \cite{Bag:98}}
From this property, we obtain the condition that $c_{k}({\bf x},
{\bf z})$ be translation invariant
\begin{eqnarray}
\label{eq:trans_dressing}
\partial^i_z c_k ({\bf x},{\bf z})=
-\partial^i_x c_k ({\bf x},{\bf z}),
\end{eqnarray} 
i.e., 
\begin{eqnarray}
\label{eq:trans_dressing'}
c_k({\bf x},{\bf z})=c_k({\bf x}-{\bf z}).
\end{eqnarray}
This results would be
also easily anticipated ones if the base fields $\phi$ and $A_i$ are
translationally 
invariant as well. But this is not always necessary to derive
(\ref{eq:trans_dressing'}). Actually in this case the base fields are not
translationally invariant under the generators of
(\ref{eq:Poincare_gen})\footnote{These unconventional coordinate
transformations on fields can be understood as the ``gauge-covariant''
coordinate 
transformation \cite{Jac:78,Jac:95} $\bar{\delta}_f A_{\mu} \equiv
{\cal L}_f A_{\mu} -\partial_{\mu}(f^{\alpha} A_{\alpha})=f^{\alpha}
F_{\alpha \mu},~ \bar{\delta}_f \phi \equiv {\cal L}_f \phi - i
f^{\alpha}A_{\alpha} \phi=f^{\alpha} (\partial_{\alpha} + i
A_{\alpha}) \phi $ under the coordinate transformation $\delta_f
x^{\mu} =-f^{\mu} (x)$ [$f^{\mu}$=constant for space-time translation,
$f^{\mu}={\omega^{\mu}}_{\nu}
x^{\nu}~({\omega^{\mu}}_{\nu}=-{\omega_{\nu}}^{\mu})$ for the space
rotation and Lorentz boost], where ${\cal L}_f$ denotes the Lie
derivative; the discrepancies about $A_0$ transformation can be traced
back to the (strong) implementation of the constraint $\pi_0 \approx 0$ which
is the gauge transformation generator of $A_0$. By recovering $\pi_0$
in the Poincar\'e generator complete equivalence will be
obtained. These results are consistent with the fact that, by
introducing the (first-class) constraints terms additionally into the
Poincar\'e generator, one can obtain the correct coordinate
transformations as is shown in Appendix {\bf A}: The additional
constraints terms compensate the gauge transformation part with gauge
function $f^{\alpha} A_{\alpha}$ in the gauge-covariant coordinate
transformation $\bar{\delta}_f$ and the correct transformations
$\delta_f A_{\mu}={\cal L}_f A_{\mu},~ \delta_f \phi ={\cal L}_f \phi$
are obtained. We thank Prof. R. Jackiw for suggesting this way of
understanding.}, 
\begin{eqnarray*}
&&\{\phi ({\bf x}), P^j _s \} =D ^j  \phi({\bf x}),
\nonumber \\
&&\{A_i ({\bf x}), P^j _s \} =-\epsilon_{ij} \frac{1}{\kappa} J_0({\bf
  x}) \approx {F^j}_i ({\bf x}), \nonumber \\
&&\{A_0 ({\bf x}), P^j _s \} =0,
\end{eqnarray*}
but the translationally
non-invariant terms cancel each others in the gauge invariant
combination (\ref{eq:phys_variable}). Furthermore, the 
condition (\ref{eq:trans_dressing}) or (\ref{eq:trans_dressing'}) also
guarantees the correct spatial translation law for all 
other gauge invariant fields in ${\cal F}^{\alpha}$\footnote{Correct
transformation for ${\cal A}_0$ also can be obtained by reviving the
$\pi_0$ term in the Poincar\'e generators similar to what was noted in
the footnote `3'. See 
Appendix {\bf A} for detail.}:
\begin{eqnarray*}
\{ {{{\cal  F}}_{\alpha}}({\bf x}), { P}^j_{s} \}= \partial^j
{{{\cal  F}}_{\alpha}}({\bf x}) ,~ {{\cal  F}}_{\alpha}= ({{\cal A}_i},
{\hat{\phi}},{{\hat{\phi}}^*}).
\end{eqnarray*}

Similarly, by considering  the time translation
\begin{eqnarray*}
\{ {\hat{\phi}}({\bf x}),{P}^{0}_{s}\}
&=&\partial ^0 \phi ({\bf x}) e^{i W ({\bf x})} + i\phi ({\bf x})  e^{i W
  ({\bf x})}\int d^2 {\bf z} ~ c_k ({\bf x}-{\bf z}) \partial^0 _z A^k
({\bf z}) \nonumber \\ 
&=& \partial^0 {\hat{\phi}}({\bf x}) -i {\hat{\phi}}({\bf x})
\int d^2{\bf z} ~\partial^0 \left( c_k ({\bf x}-{\bf z})\right) {A}^k
({\bf z}) 
\end{eqnarray*}
we obtain the correct time translation if the boundary integral
$
\int d^2 {\bf z} ~ \partial^k_z \left(c_k({\bf x}-{\bf z}) 
{ A}^0({\bf z})\right) 
$
vanishes and $c_k({\bf x}-{\bf z})$ be time
independent.  This property is also satisfied for all other gauge
invariant fields and the
results read in a compact form
\begin{eqnarray*}
\{ {{{\cal  F}}_{\alpha}}({\bf x}), { P}^0_{s} \}= \partial^0
{{{\cal F}}_{\alpha}}({\bf x}).
\end{eqnarray*}
Contrast to this, the transformation for the base fields are not
correctly generated, {i.e., }
\begin{eqnarray*}
&&\{\phi ({\bf x}), P^0_s \} =D_0 \phi ({\bf x}),
\nonumber \\
&&\{A_i ({\bf x}), P^0 _s \} =-\epsilon_{ij} \frac{1}{\kappa} J_j
({\bf x})={F^0}_i ({\bf x}), \nonumber \\
&&\{A_0 ({\bf x}), P^0 _s \} =0.
\end{eqnarray*}

\begin{center}
{\bf 2. Space rotations and Lorentz boost }
\end{center}

For the space rotation and Lorentz boost, contrast to the
translations, the anomalies are present even in the transformation for
${\cal F}^{\alpha}$ since in that case they represent the spin or
other properties of ${\cal F}_{\alpha}$. The brackets with the rotation
generator are expressed as
\begin{eqnarray*}
     \{ \hat{\phi}({\bf x}), M^{12}_s \}
 &=&\epsilon_{ij} x_i (\partial_j + i A_j ) \phi ({\bf x}) e^{i W
 ({\bf x})} + i \phi({\bf x}) e^{i W ({\bf x})} \int d^2 {\bf z} ~ c_k
 ({\bf x}-{\bf 
 z}) z^k \frac{1}{\kappa} J_0 ({\bf z}) \nonumber \\
&=& \epsilon_{ij} x_i \partial_j \hat{\phi}({\bf x}) -i
       \Xi^{12}({\bf x}) \hat{\phi}({\bf x}), \nonumber \\
\{ {\cal A}_i ({\bf x}), M^{12}_s \}
&=&\epsilon_{jk} x_j \partial_k  A_i ({\bf x}) -\epsilon_{ij} A_j ({\bf x}) -
 \partial ^x _i \int d^2 {\bf z} ~ c_k  ({\bf x}-{\bf
 z}) \left[ \epsilon^{jl}  z^j \partial^l _z A^k ({\bf z}) -\epsilon_{kj}
 A^j ({\bf z}) \right] \nonumber \\
&=& \epsilon_{jk} x_j \partial_k  {\cal A}_i ({\bf x}) -\epsilon_{ij}
 {\cal A}_j ({\bf x})+ \partial_i \Xi ^{12}({\bf x}), 
\end{eqnarray*}
and the brackets with the Lorentz boost are
\begin{eqnarray*}
     \{ \hat{\phi}({\bf x}), M^{0j}_s \}
&=& ( x^0 \partial^j - x^j \partial^0) \phi ({\bf x}) e^{i W
 ({\bf x})}  \nonumber \\
 && ~~~+ i \phi({\bf x}) e^{i W ({\bf x})} \int d^2 {\bf z} ~ c_k
 ({\bf x}-{\bf 
 z}) \left[(z^0 \partial ^j _z -z^j \partial ^0) A^k ({\bf z})+ \delta_{kj}
 A_0 ({\bf z})\right] \nonumber \\
&=&  x^0 \partial^j \hat{\phi}({\bf x})  
 - x^j \partial^0 \hat{\phi} ({\bf x}) -i \Xi ^{0j}({\bf x})\hat{\phi}
 ({\bf x}),  \nonumber \\ 
\{ {\cal A}_i({\bf x}), M^{0j}_s \}
 &=&( x^0 \partial^j - x^j \partial^0) A_i ({\bf x}) -\delta_{ij} A_0
 \nonumber \\
&& ~~~-\partial^x _i \int d^2 {\bf z} ~ c_k  ({\bf x}-{\bf
 z}) \left[(z^0 \partial ^j _z -z^j \partial ^0) A^k ({\bf z})+ \delta_{kj}
 A_0 ({\bf z})\right] \nonumber \\
&=&  x^0 \partial^j {\cal A}_i({\bf x})  - x^j \partial^0 {\cal
  A}_i({\bf x})-\delta_{ij} {\cal A}_0 + \partial_i  \Xi 
 ^{0j}({\bf x}) \phi ({\bf x});  
\end{eqnarray*}
or in the compact form these are expressed as follows
\begin{eqnarray}
\label{eq:ano_transf}
\{ {\cal F}_{\alpha}({\bf x}), M^{\mu \nu}_s \}&=& 
   x^{\mu} \partial^{\nu}{\cal F}_{\alpha}({\bf x})
   -x^{\nu} \partial^{\mu} {\cal F}_{\alpha}({\bf x}) 
   +\Sigma^{\mu \nu}_{\alpha \beta} {\cal F}_{\beta}({\bf x})
   +\Omega^{\mu \nu}_{\alpha}({\bf x}), \nonumber \\
\Omega^{\mu \nu}_{\hat{\phi} }({\bf x})&=&
   -i {\Xi}^{\mu \nu} ({\bf x}) 
\hat{\phi} ({\bf x}), \nonumber \\
\Omega^{\mu \nu}_{ \hat{\phi}^{*} }({\bf x})&=&
   i {\Xi}^{\mu \nu} ({\bf x}) 
\hat{\phi}^* ({\bf x}), \nonumber \\
\Omega^{\mu \nu}_{{\cal A}_i}({\bf x})&=&
  \partial_i {\Xi}^{\mu \nu} ({\bf x}),
\end{eqnarray}
where
\begin{eqnarray*}
&&\Xi ^{\mu \nu} =- \Xi^{\nu \mu}, \nonumber \\
&&{\Xi}^{12}({\bf x})= \epsilon_{ij} x_{i} {\cal A}^{j} ({\bf x})
+ \frac{1}{\kappa}
 \int d^2 {\bf z} ~ z_{k} c_{k}({\bf x}-{\bf z}) J_0({\bf z}),\nonumber \\
&&{\Xi}^{0i}({\bf x})=-x_i {\cal A}^0({\bf x})
-\frac{1}{\kappa} \int d^2 {\bf z} ~ z_i 
  \epsilon _{kj} c_{k}({\bf x}-{\bf z}) J^{j}({\bf z})
\end{eqnarray*}
with the spin-factors $\Sigma ^{\mu \nu}_{\alpha \beta}= \eta ^{\mu
  \alpha} \eta ^{\nu}_{\beta} -\eta^{\mu}_{\beta} \eta^{\nu \alpha},
  ~\Sigma ^{\mu \nu}_{\phi (\phi^*)}=0 $
for the gauge and scalar fields, respectively.
As a comparison, the corresponding brackets for the base fields are as
  follows: 
\begin{eqnarray*}
&&\{\phi ({\bf x}), M^{12}_s \} =\epsilon_{ij} x^i D ^j \phi ({\bf x}),
\nonumber \\
&&\{A_i ({\bf x}), M^{12}_s \} =x_i \frac{1}{\kappa} J_0 ({\bf x})\approx x_i
\epsilon_{kl} \partial _k A^l ({\bf x}), \nonumber \\
&&\{A_0 ({\bf x}), M^{12}_s \} =0,
\end{eqnarray*}
and
\begin{eqnarray*}
\label{eq:ano_transf_bare}
&&\{\phi ({\bf x}), M^{0j}_s \} =(x^0 D^j - x^j D^0)
\phi ({\bf x}), 
\nonumber \\
&&\{A_i ({\bf x}), M^{0j}_s \} =x^0 {F^j}_i ({\bf x}) -x^j F_{i0}
({\bf x}), \nonumber \\ 
&&\{A_0 ({\bf x}), M^{0j}_s \} =0
\end{eqnarray*}
for the space rotation and Lorentz transformations, respectively, which show
the incorrect transformation for the gauge {\it varying} base fields;
we must supplement the constraint term $\Gamma$ in order to give the correct
transformation even for the gauge varying base fields as well as the
gauge invariant fields ${\cal F}_{\alpha}$. 
The anomalous term $\Omega^{\mu \nu}_{\alpha}$ for the transformation
of ${\cal F}_{\alpha}$ in (\ref{eq:ano_transf}) is gauge invariant
because it is 
expressed only with the ${\cal F}_{\alpha}$.  At
first, it seems odd that the gauge invariant variables do have the
anomaly, but as will be clear in later section, these variables are
nothing but 
the Hagen's rotational anomaly term and other gauge restoring terms in
GFF. This will be treated in Sec. IV. But here, it will be interesting
to note that ${\cal A}_{\mu}$ can be re-expressed completely by the
matter currents as 
\begin{eqnarray}
\label{eq:AA_i_solution}
{\cal A}_{i} ({\bf x}) &=&A_i ({\bf x}) - \int d^2 {\bf z} ~\partial ^x _i
c_k({\bf x}-{\bf z}) A^k ({\bf z}) \nonumber \\
&=&A_i ({\bf x}) - \int d^2 {\bf z} 
~c_k({\bf x}-{\bf z}) \left( F_{ki}({\bf z}) - \partial _k A_i ({\bf
    z}) \right)
\nonumber \\
&=& - \int d^2 {\bf z}~ c_k ({\bf x}-{\bf z}) F_{ki} ({\bf z})
\nonumber \\     
&\approx& -\frac{1}{\kappa} \int d^2 {\bf z} ~ \epsilon _{ik}
 c_{k} ({\bf x}-{\bf z}) J^0 ({\bf z}), \\
\label{eq:AA_0_solution}
{\cal A}_{0} ({\bf x}) &=&A_0 ({\bf x}) - \int d^2 {\bf z} ~ 
c_k({\bf x}-{\bf z})\partial_0 A^k ({\bf z}) \nonumber \\
&=&A_0 ({\bf x}) - \int d^2 {\bf z} ~ 
c_k({\bf x}-{\bf z}) \left(F_{k0}({\bf z}) + \partial ^k A^0 ({\bf z})\right)
\nonumber \\ 
&=&- \int d^2 {\bf z}~ c_k ({\bf x}-{\bf z}) F_{k0} ({\bf z}) \nonumber \\ 
&=&-\frac{1}{\kappa} \int d^2 {\bf z} ~ \epsilon _{kj} c_{k} 
({\bf x}-{\bf z}) J^j ({\bf z}) 
\end{eqnarray} 
: The third lines in (\ref{eq:AA_i_solution}) and
(\ref{eq:AA_0_solution}) are just the results of integration by parts without
recourse to the particular properties of CS theory; in the last steps
we used the constraint $T \approx 0$ 
and the Euler-Lagrange equation of 
(\ref{eq:CS_action}), 
$
F^{0k} =-\frac{1}{\kappa} \epsilon_{kj} J^j   
$
, respectively, 
which are genuine to the CS theory. It is
remarkable that the expressions (\ref{eq:AA_i_solution}) and
(\ref{eq:AA_0_solution}) in terms of $J^{\mu}$, 
which solves the constraint and equation of motion, were obtained
without solving the differential equations but from the simple
algebraic manipulation by imposing the constraint and equation of
motion. 
Moreover, the
anomalous terms are, then, expressed as 
\begin{eqnarray}
\label{eq:ano_factor}
  \Xi^{12} &=& \frac{1}{\kappa} \int d^2 {\bf z} ~(x^k -z^k) c_k ({\bf
  x}-{\bf z}) 
  J_0 ({\bf z}) \nonumber \\
  \Xi^{0i} &=& -\frac{1}{\kappa} \int d^2 {\bf z} ~(x^i -z^i) c_k
  ({\bf x}-{\bf z}) 
  \epsilon_{kj} J^j ({\bf z}).
\end{eqnarray}

These solutions are
similar to the Coulomb gauge solution \cite{Hag:84}
and hence imply the
similarity of GIF to GFF with the Coulomb gauge in
particular.\footnote{This has the 
  same origin to what has been observed in a different context in
  Ref. \cite{Fad:88}
} (This will be
discussed again in Sec. IV. {\bf A} in a different context.)
 However it should be
noted that the Lorentz anomaly does not occur in the transformation of
the current $J^{\mu}$, even though ${\cal A}_{\mu}$, which is
expressed by $J^{\mu}$ as given
above, does have the anomaly: This will be connected to the fact that
the anomaly depends on the dressing $ c_k ({\bf x}-{\bf z})$ but
$J^{\mu}$ is already gauge invariant without recourse to that dressing.

\begin{center}
{ \bf III. Quantization} 
\end{center}

The quantization in our gauge invariant formulation is carried out by
assuming the (equal time) quantum commutation relation
\begin{eqnarray}
\label{eq:quantization}
&&[ A^i_{op}({\bf x}), A^j_{op}({\bf y}) ] =\frac{i \hbar}{\kappa}
 \epsilon^{ij} 
 \delta^{2} ({\bf x}-{\bf y}), \nonumber \\
&&[ \phi _{op}({\bf x}), \pi _{op}({\bf y}) ]=
 [\phi^{\dagger} _{op} ({\bf x}), \pi^{\dagger} _{op} ({\bf y}) ] = 
i \hbar \delta^{2}({\bf x}-{\bf y}), \\
&&\mbox{others~vanish} \nonumber
\end{eqnarray}
for the operator valued fields $A^i _{op}, \phi_{op}, \pi_{op}=(D_0
\phi )^{\dagger}_{op}$, and their complex conjugates, and the
physical states $\left| \Psi_{\mbox{phys}}\right>$, which are
annihilated by the Gauss' law constraint (\ref{eq:T}) (with normal
ordering : :) 
\begin{eqnarray}
\label{eq:Gauss_law}
T_{op} \left| \Psi_{\mbox{phys}} \right>  =0. 
\end{eqnarray}
Here, we note that the quantum commutation relations in
(\ref{eq:quantization}) are not gauge 
independent because the involved field operators $A^i _{op}, \phi_{op}$,
... etc. are gauge {\it varying} ones. 
So, we first consider the
commutation (or exchange) relations for the gauge invariant variables
${\cal F}^{\alpha}_{op}$ which are more basic objects in our formulation.  

\begin{center}
{\bf A. Operator exchange relations}
\end{center}

At the classical level, the basic brackets
between the gauge invariant canonical fields ${\cal F}^{\alpha}$
become as follows 
\begin{eqnarray}
\label{eq:graded_bracket}
\{ {\hat{\phi}}({\bf x}),{\hat{\phi}}({\bf y}) \} &=&
  -{\hat{\phi}}({\bf x}){\hat{\phi}}({\bf y}) \frac{1}{\kappa}
  \Delta({\bf x}-{\bf y}), \nonumber \\
\{ {\hat{\phi}}({\bf x}),{\hat{\phi}}^{*}({\bf y}) \} &=&
  {\hat{\phi}}({\bf x}){\hat{\phi}}^* ({\bf y}) \frac{1}{\kappa}
  \Delta({\bf x}-{\bf y}), \nonumber \\
\{ {\hat{\phi}}({\bf x}),{\hat{\pi}}({\bf y}) \} &=&
  \delta^2 ({\bf x}-{\bf y}) + {\hat{\phi}}({\bf x}){\hat{\pi}}({\bf
  y}) \frac{1}{\kappa} 
  \Delta({\bf x}-{\bf y}), \nonumber \\
\{ {\hat{\phi}}({\bf x}),{\hat{\pi}}^{*}({\bf y}) \} &=&
  -{\hat{\phi}}({\bf x}){\hat{\pi}}^* ({\bf y}) \frac{1}{\kappa}
  \Delta({\bf x}-{\bf y}), \nonumber \\
\{ {\hat{\pi}}({\bf x}),{\hat{\pi}}({\bf y}) \} &=&
  -{\hat{\pi}}({\bf x}){\hat{\pi}}({\bf y}) \frac{1}{\kappa}
  \Delta({\bf x}-{\bf y}), \nonumber \\
\{ {\hat{\pi}}({\bf x}),{\hat{\pi}}^{*}({\bf y}) \} &=&
  {\hat{\pi}}({\bf x}){\hat{\pi}}^* ({\bf y}) \frac{1}{\kappa}
  \Delta({\bf x}-{\bf y}), \nonumber \\
\{ {\cal A}_i({\bf x}),{\cal A}_j({\bf y}) \} &=&
  \frac{1}{\kappa} \left[\epsilon_{ij}\delta^2({\bf x}-{\bf y})
  +\xi_{ij}({\bf x}-{\bf y}) + \partial_i^x \partial_j^y 
\Delta({\bf x}-{\bf y}) \right], \nonumber \\
\{{\cal A}_i({\bf x}), \hat{\phi}({\bf y}) \} &=&
  -\frac{i}{\kappa}{\hat{\phi}}({\bf y}) \left[ \epsilon_{ik}c_k({\bf y}-
  {\bf x})
   + \partial^x_i \Delta({\bf x}-{\bf y}) \right], \nonumber \\
\{{\cal A}_i({\bf x}), \hat{\pi}({\bf y}) \} &=&
  \frac{i}{\kappa}{\hat{\pi}}({\bf y}) \left[ \epsilon_{ik}c_k({\bf y}-
  {\bf x})
   + \partial^x_i \Delta({\bf x}-{\bf y}) \right].
\end{eqnarray}
Here, we have introduced two functions
\begin{eqnarray}
\label{eq:Delta}
&&\Delta({\bf x}-{\bf y})=\int d^2 {\bf z}~ \epsilon^{kl} c_k({\bf x}-{\bf
  z}) c_l({\bf y}-{\bf z}) , \\
\label{eq:Xi}
&&\xi_{ij}({\bf x}-{\bf
  y})=\epsilon_{ik}\partial^y_j c_k({\bf y}-{\bf x}) -
\epsilon_{jk}\partial^x_i c_k({\bf x}-{\bf y}),
\end{eqnarray} 
which are totally antisymmetric under the interchange
  of all the indices, {i.e.,} 
\begin{eqnarray}
\label{eq:Delta_sym}
 \Delta({\bf x}-{\bf y})= -\Delta({\bf y}-{\bf x}), \\
\label{eq:Xi_sym}
 \xi_{ij}({\bf x}-{\bf y})=-\xi_{ji}({\bf y}-{\bf x}).
\end{eqnarray}
Then, the corresponding quantum commutation (exchange) algebras are
\begin{eqnarray}
\label{eq:graded_comm}
\hat{\phi}_{op}({\bf x})\hat{\phi}_{op}({\bf y})   &=&
  \hat{\phi}_{op}({\bf y})\hat{\phi}_{op}({\bf x}) e^{- \frac{i
  \hbar}{\kappa} 
  \Delta ({\bf x}-{\bf y})}, \nonumber \\
 \hat{\phi}_{op} ({\bf x}) \hat{\phi}^{\dagger}_{op}({\bf y})  &=&
  \hat{\phi}^{\dagger}_{op}({\bf y})\hat{\phi}_{op} ({\bf x})
  e^{\frac{i \hbar}{\kappa}
  \Delta ({\bf x}-{\bf y})}, \nonumber \\
\hat{\phi}_{op}({\bf x})\hat{\pi}_{op}({\bf y})   &=& \delta^2({\bf
  x}-{\bf y}) 
  + \hat{\pi}_{op}({\bf y})\hat{\phi}_{op}({\bf x}) e^{\frac{i \hbar}{\kappa}
  \Delta ({\bf x}-{\bf y})}, \nonumber \\
 \hat{\phi}_{op} ({\bf x}) \hat{\pi}^{\dagger}_{op}({\bf y})  &=&
  \hat{\pi}^{\dagger}_{op}({\bf y})\hat{\phi}_{op} ({\bf x})
  e^{-\frac{ i \hbar}{\kappa}
  \Delta ({\bf x}-{\bf y})}, \nonumber \\
\hat{\pi}_{op}({\bf x})\hat{\pi}_{op}({\bf y})   &=&
  \hat{\pi}_{op}({\bf y})\hat{\pi}_{op}({\bf x}) e^{- \frac{i \hbar}{\kappa}
  \Delta ({\bf x}-{\bf y})}, \nonumber \\
 \hat{\pi}_{op} ({\bf x}) \hat{\pi}^{\dagger}_{op}({\bf y})  &=&
  \hat{\pi}^{\dagger}_{op}({\bf y})\hat{\pi}_{op} ({\bf x})
  e^{\frac{i \hbar}{\kappa}
  \Delta ({\bf x}-{\bf y})}, \nonumber \\
\left[ {{\cal A}_i}_{op}({\bf x}),{{\cal A}_j}_{op}({\bf y}) \right] &=&
  \frac{i \hbar}{\kappa} \left[\epsilon_{ij}\delta^2({\bf x}-{\bf y})
  +\xi_{ij}({\bf x}-{\bf y}) + \partial_i^x \partial_j^y 
\Delta ({\bf x}-{\bf y}) \right], \nonumber \\
\left[ {{\cal A}_i}_{op}({\bf x}), \hat{\phi}_{op}({\bf y}) \right]
  &=&-\frac{\hbar}{\kappa} \hat{\phi}({\bf y}) \left
  [ \epsilon_{ik}c_k({\bf y}- 
  {\bf x})
   + \partial^x_i \Delta({\bf x}-{\bf y}) \right], \nonumber \\ 
\left[ {{\cal A}_i}_{op}({\bf x}), \hat{\pi}_{op}({\bf y}) \right]  &=&
  \frac{\hbar}{\kappa} \hat{\pi}({\bf y}) \left[ \epsilon_{ik}c_k({\bf y}-
  {\bf x})
   + \partial^x_i \Delta({\bf x}-{\bf y}) \right],
\end{eqnarray}
where 
\begin{eqnarray}
\label{eq:op_phys_variable}
&&\hat{\phi}_{op}({\bf x}) \equiv \phi _{op}({\bf x})
exp\left[iW_{op}({\bf x})\right], \nonumber \\
&&\hat{\pi}_{op}({\bf x}) \equiv \pi_{op}({\bf x})
exp\left[-iW_{op}({\bf x})\right], \nonumber \\
&&{{\cal A}_{\mu}}_{op}({\bf x})\equiv {A_{\mu}}_{op}({\bf
  x})-\partial_{\mu}W_{op} ({\bf x}) 
\end{eqnarray}
with $W_{op} ({\bf x}) = \int  d^2 {\bf z} ~c_k ({\bf x}-{\bf z}) A^k
_{op} ({\bf 
z})$ and we have used the formula 
$
  e^A e^B = e^{A+B+\frac{1}{2}[A,B]}$ with $[[A,B],A]=[[A,B],B]=0. 
$
Here, we note that the path-ordering is not needed in defining the
exponential factor of (\ref{eq:op_phys_variable}) although
$A^i_{op}$'s do not commute by 
themselves: They are non-commuting only for the
same positions such that $A^{i}_{op}$ of the adjacent points of the
integration range commutes: Only for the non-Abelian case, the
path-ordering is needed\footnote{
If we consider the situation where are the crossings of the
  contours in the line integral representation of $W({\bf x})$, the
  path-ordering is required \cite{Dun:89}; See also footnote `21'  in
  this paper for this problem.}. These results look like the graded
commutation 
relations of the {\it anyon field} \cite{Wil:82} but it is found that this is
not the case {\it always} \cite{Hag:84,Hag:89}
. This will be treated in the Sec. V in detail.

\begin{center}
{\bf B. Physical states : Algebraic construction}
\end{center}

There is well-known way to construct the physical states:
If the vacuum state $\left| 0 \right>$ is a physical state which
satisfies (\ref{eq:Gauss_law}), the state
$
{\cal O} ({{\cal A}_i}_{op}, \hat{\phi}_{op},
\hat{\phi}_{op}^{\dagger}) \left| 0 \right >
$, which is the power series function ${\cal O}$, is also a physical state
[ this is because $T {\cal O} \left| 0 \right> = ([T, {\cal O} ] +
{\cal O} T ) \left| 0 \right> = [T, {\cal O}] \left| 0 \right > =0$ is
satisfied if ${\cal O}$ is gauge-invariant].
As a simplest case, let us consider the state
\begin{eqnarray}
\label{eq:1_particle_state}
\hat{\phi}_{op}({\bf x}) \left| 0 \right>.
\end{eqnarray}
Then, it is found that this state has one unit charge at the
operator position ${\bf x}$ of the operator in addition to the vacuum charge
\begin{eqnarray*}
J_0 ({\bf y}) \hat{\phi}_{op}({\bf x}) \left| 0 \right> &=&
 \left\{ [J_0 ({\bf y}),\hat{\phi}_{op}({\bf x})] +\hat{\phi}_{op}({\bf
 x})J_0 ({\bf y}) \right\}  \left| 0 \right> \nonumber \\
&=&\left[ \delta({\bf x}-{\bf y}) + \bar{J_0} ({\bf y})
 \right]\hat{\phi}_{op}({\bf x}) \left| 0 \right>,  
\end{eqnarray*}
where we have used the commutation relation
$
[J_0 ({\bf y}),
\hat{\phi}_{op}({\bf x}) ]=\delta^2({\bf x}-{\bf y})
\hat{\phi}_{op}({\bf x}) 
$
and $\bar{J_0} ({\bf y})$ is the charge density of the vacuum at the
position ${\bf y}$. The state (\ref{eq:1_particle_state}) has also the
gauge field cloud 
around the position ${\bf x}$ of $\hat{\phi}_{op}({\bf x})$ as 
\begin{eqnarray}
\label{eq:A_i_phi}
A^i_{op} ({\bf y}) \hat{\phi}_{op}({\bf x}) \left| 0 \right>&=&
 \left\{ [A^i_{op} ({\bf y}),\hat{\phi}_{op}({\bf x})] +\hat{\phi}_{op}({\bf
 x})A^i_{op} ({\bf y}) \right\}  \left| 0 \right> \nonumber \\
&=&\left[ \frac{\hbar }{\kappa} \epsilon_{ik} c_k ({\bf x}-{\bf y}) +
\bar{A^i} ({\bf y})\right]\hat{\phi}_{op}({\bf x}) \left| 0 \right>, \\
B_{op} ({\bf y}) \hat{\phi}_{op}({\bf x}) \left| 0 \right>&=&
 \epsilon_{ij} \partial^y _i \left(A^j_{op} ({\bf y})\hat{\phi}_{op}({\bf
 x})\right)  \left| 0 \right> \nonumber \\
&=&\left[ \frac{\hbar}{\kappa} \delta^2 ({\bf x}-{\bf y}) +
\bar{B} ({\bf y}) \right]\hat{\phi}_{op}({\bf x}) \left| 0 \right>,
\end{eqnarray}
where $\bar{A^i} ({\bf y}),~\bar{B} ({\bf y})$ are the fields of the
vacuum at the point ${\bf y}$ and we have used the commutation relation, 
$
[{A^i}_{op} ({\bf y}), \hat{\phi}_{op}({\bf x})
]=\frac{\hbar}{\kappa} \hat{\phi}_{op}({\bf x})\epsilon_{ik} c_{k}
({\bf x}-{\bf y}) 
$
in (\ref{eq:A_i_phi}): The state $\hat{\phi}_{op}({\bf x}) \left| 0
\right>$ has the 
gauge varying vector field 
$
{ a}^i ({\bf y})=\frac{\hbar}{\kappa} \epsilon_{ik}
c_{k} ({\bf x}-{\bf y}),
$ 
as well as the gauge invariant point magnetic field 
$
{b}({\bf y}) (=\epsilon_{ij}\partial_i
a^j)=\frac{\hbar}{\kappa}\delta^2({\bf y}-{\bf x}).
$
This is understood as that the $\hat{\phi}_{op}({\bf x })$ creates one
charged 
particle at the position ${\bf x}$ together with the vector field $a^i
({\bf y})$ and the point 
magnetic flux $\int d^2 {\bf y}~ b ({\bf
  y})=\frac{\hbar}{\kappa}$. Therefore, let us call
$\hat{\phi}_{op}$ the charge-flux composite (CFC) operator. On the
other hand, $\phi
_{op}({\bf x })$ creates one
charged particle at ${\bf x}$ without gauge field cloud, and $e^{i W _{op}
  ({\bf x})}$ creates the point magnetic flux at ${\bf x}$ without charge.
[This situation is
in contrast to the QED case where $\hat{\phi}_{op}({\bf x})$ creates
the gauge invariant (physical) electron together with only the gauge
invariant electric field \cite{Dir:55,Lav:93}.] Similarly,
$\hat{\phi}^{\dagger} _{op}$ 
annihilates one CFC operator and let us call it the anti-CFC one. Now,
$N$ CFC state becomes
\begin{eqnarray}
\label{eq:CFC_op}
  \sum ^L _{a=1} c_a \prod ^a_{i=1}  \hat{\phi}_{op} ({\bf x}_i) \left|
  0 \right>
\end{eqnarray}
with the c-number coefficient $c_a$. Furthermore, more general state
with $N$ CFC 
and $M$ anti-CFC is expressed as
$
 \sum ^M _{b=1}  \sum ^L _{a=1} c_{ab} 
  \prod ^b_{j=1} \hat{\phi}_{op} ({\bf x}_j) \prod ^a_{i=1}
 \hat{\phi}_{op}^{\dagger} ({\bf x}_i)\left| 0 \right>
$
when ${\hat{\phi} _{op}}^{\dagger}$'s are placed to the right-hand side of
${\hat{\phi} _{op}}$'s. In the case of (\ref{eq:CFC_op}), the CFC
operators can be 
re-arranged as a factorized form
\begin{eqnarray*}
  &&\prod^L _{j=1} e^{i W_{op} ({\bf x}_{j})} \prod ^L _{i=1}
 \phi_{op} ({\bf x}_{i}) 
 \left| 0 \right> 
 =e^{-\frac{1}{2} \frac{ i \hbar}{\kappa} \sum^{L-1} _{i=1} \Delta
 ({\bf x}_{i}-{\bf x}_{i+1}) }~ e^{i \sum ^L_{i=1} W_{op} ({\bf x}_{i})
 } \prod ^L _{i=1}  \phi_{op} ({\bf x}_{i})
 \left| 0 \right> 
\end{eqnarray*}
in an appropriate order of $\hat{\phi}_{op}$'s and the first
exponential factor will show the multi-valuedness under the exchanging
any two CFC's if $\Delta ({\bf x}_{i}-{\bf x}_{i+1})$ does not vanish. 

Let me now consider finally the state
$
  {\cal A}^i _{op} ({\bf x}) \left| 0 \right >.
$
Then, it is easy to see that this state does not carry the charge nor
magnetic flux in addition to the vacuum charge
\begin{eqnarray*}
  &&J_0 ({\bf y}){\cal A}^i _{op} ({\bf x}) \left| 0 \right >=\bar{J_0}
  ({\bf y}) {\cal A}^i _{op} ({\bf x}) \left| 0 \right >, \nonumber \\
  &&B ({\bf y}){\cal A}^i _{op} ({\bf x}) \left| 0 \right >=\bar{B}
  ({\bf y}) {\cal A}^i _{op} ({\bf x}) \left| 0 \right >
\end{eqnarray*}
from the communication relations
$
  [ J_0 ({\bf y}), {\cal A}^i _{op} ({\bf x}) ] =0,  
 [ B ({\bf y}), {\cal A}^i _{op} ({\bf x}) ] =0.
$
But, it just carries the gauge varying vector field
$
  a^j ({\bf y}) = - \frac{i \hbar}{\kappa} \left[ \epsilon_{ij}
  \delta^2 ({\bf x}- {\bf y}) + \epsilon_{kj} \partial_i^x c_k ({\bf
  x}-{\bf y}) \right]
$
in addition to that of the vacuum:
  \begin{eqnarray*}
    A^j _{op} ({\bf y}) {\cal A}^i _{op} ({\bf x}) \left| 0 \right >=
\left\{ \frac{- i \hbar}{\kappa} \left[\epsilon_{ij} \delta^2 ({\bf
      x}-{\bf y}) 
+ \epsilon_{kj} \partial_i^x c_k ({\bf x} -{\bf y}) \right] +
\bar{A}^j  ({\bf y}){\cal A}^i _{op} ({\bf x}) \right\}\left| 0 \right >. 
  \end{eqnarray*}
The most general state, then, will be
\begin{eqnarray}
\label{eq:phys_wave_1} 
\sum^N_{c=1} \sum ^M _{b=1}  \sum ^L _{a=1} c_{cba}
   \prod ^c_{k=1} {\cal A}^k _{op}
  ({\bf x}_k)  \prod ^b_{j=1} \hat{\phi}_{op} ({\bf x}_j)  \prod^a_i
  \hat{\phi}_{op}^{\dagger} ({\bf x}_i)\left| 0 \right>
\end{eqnarray}
with the c-number coefficient $c_{cba}$. Here, we note that there
seems to be no general
reason to omit the purely gauge field ${\cal A}^k_{op}$ part in this
construction. However, that part is physically doubtful because it can
imply the independent gauge field dynamics contrast to the nature of
the CS gauge field. Moreover, the representation of the physical
states are not unique: It depends on what gauge invariant operators are
fundamental and the physical states with different representations are
not equivalent in general.  As a more explicit approach, which fixes
these problems and shows the more detailed form of the states, we consider
the functional Schr\"odinger picture approach.
\begin{center}
 {\bf C. Physical wavefunctional in Schr\"ondinger picture }
\end{center}

In the previous section {\bf B}, we have considered a general
algebraic construction for the physical states
$\left|\Psi_{\mbox{phys}} \right>$. In this 
section, we consider a more explicit way to construct them, especially
in the Schr\"odinger picture \cite{Sch_picture_rev}.

To go to the Schr\"odinger picture, we must choose a representation
for the field commutation relation (\ref{eq:quantization}). Instead of
taking the 
rotationally non-symmetric representations, which take one (spatial)
component of $A_i$ as the conjugate momenta of the other component
\cite{Dun:89}, we take the rotationally symmetric representation which
shows the contents of the gauge invariant operators more explicitly, as
in the previous section {\bf B}. 

To this end, we note that ${\cal A}_i$ can be expressed as
\begin{eqnarray}
  \label{eq:AA-decomp}
  {\cal A}_i =\epsilon_{ij} \partial^{-1}_j B + \partial _i \eta
\end{eqnarray}
which solves the equation
$
  \epsilon_{ij} \partial_i {\cal A}^j =\epsilon_{ij} \partial_i A^j
  =B
$
with a {\it gauge invariant} scalar field $\eta$.  The field $\eta$ is
determined as
\begin{eqnarray}
  \label{eq:eta}
  \eta({\bf x}) =-\nabla^{-2} \left[ \partial^i A_i ({\bf x}) +
  \nabla^2 W ({\bf x}) \right]
\end{eqnarray}
by considering the divergence (\ref{eq:div_AA})
\begin{eqnarray*}
  \partial^i {\cal A}_i ({\bf x})&=&\partial^i {A}_i ({\bf x})
  +\nabla^2  W ({\bf x}) \nonumber \\
  &=& -\nabla^2 \eta ({\bf x}).
\end{eqnarray*}
Using the expressions of
  (\ref{eq:AA-decomp}) and 
  (\ref{eq:eta}), the base field $A_i$ of (\ref{eq:new_decomp}) can be
  re-arranged as follows 
  $
     A_i =\epsilon_{ij} \partial^{-1}_j B + \partial _i (\eta + W),
  $
where the first term corresponds to the transverse component $A_i^T$,
and the second term corresponds to the
longitudinal component $A^L_i$ of (\ref{eq:tl_decomp})\footnote{Here,
  one finds that all the longitudinal 
  parts are not participated in the gauge transformation but only
  $\partial_i W$ does the work}; but when the zero-mode can be
neglected, one finds that the combined quantity `$\eta +W$' becomes the
usual form of the longitudinal mode which appears in
  (\ref{eq:phys_variable_usual}): 
\begin{eqnarray}
\label{eq:eta+W}
 \eta +W &=&-\nabla^{-2} \partial^i A_i - \nabla^{-2} \nabla^2 W + W
  \nonumber \\
   &=&-\partial^i \nabla^{-2} A_i -W + W \nonumber \\
   &=&\partial_i^{-1}  A_i\nonumber \\
   &=&\partial_i^{-1}  A^L_i.
\end{eqnarray}

Now, let us consider a representation for the Schr\"odinger
picture. To this end, we first note that $W({\bf x})$ is the (unique)
canonical conjugate of $B$ while $\eta$ is completely decoupled in the
canonical conjugate sector of $B$, and furthermore $\eta$ can not be
simultaneously diagonalized with $W$ from the commutation relations
[we omit the subscript `$op$' hereafter]:
\begin{eqnarray*}
&&[ W ({\bf x}, B ({\bf y}) ] = \frac{i \hbar}{\kappa} \delta^2 ({\bf
  x}-{\bf y}), \\
  &&[\eta({\bf x}), B ({\bf y}) ] = 0, \\
  &&[W({\bf x}), \eta({\bf y}) ] = \frac{i \hbar}{\kappa} \nabla^{-2}
  \left[ \epsilon_{ik} \partial_i^y c_k ({\bf x}-{\bf y}) -\nabla^2
  \Delta ({\bf x}-{\bf y}) \right] \\
         &&~~~~~~~~~~~~~~~~\neq 0 . \nonumber 
\end{eqnarray*}
Hence, for the representation with diagonalized $W({\bf x}), B ({\bf
  y})$ will acts as a (functional) operators
\begin{eqnarray}
  \label{eq:naive_rep}
  B({\bf y}) \left|\Psi \right> \rightarrow \frac{\hbar}{i \kappa}
  \frac{\delta}{\delta W ({\bf y})} \Psi (W)
\end{eqnarray}
for any state functional $\Psi (W)$. This is a usual step that can be
performed, although the detail form of conjugate moment $B$ is different
from, theory to theories when $A_i$'s commute with themselves [see
Appendix {\bf B} for the analysis about this usual case in our
context], where $\eta$ has 
no important role in the construction of $\Psi $. Now, here one
meets an unusual situation where the naive expectation
(\ref{eq:naive_rep}) is not 
valid and $\eta$ is crucial as well as $W$ in the construction of
$\Psi $ : This results from the fact that neither $W({\bf x})$ nor
$\eta({\bf y})$ can be taken as an diagonalized base in the Scr\"odinger
picture  because of the commutation relations
\begin{eqnarray*}
  &&[W ({\bf x}), W ({\bf y}) ]=\frac{i \hbar}{\kappa} \Delta ({\bf
  x}-{\bf y}), \\
&&[\eta ({\bf x}), \eta ({\bf y}) ]=-\frac{i \hbar}{\kappa}
  \nabla^{-2} \xi_{kk}  
({\bf x}-{\bf y}) +\frac{i \hbar}{\kappa} \Delta ({\bf
  x}-{\bf y}),
\end{eqnarray*}
which all are non-vanishing in general.\footnote{ For the coincident points
  ${\bf x}={\bf y}$, one can make them vanish within a regularization
  prescription \cite{Park:94}. But, the general multi-particle states
  with the 
  distinguishable positions can not be diagonalized with respect to
  neither $W$ nor $\eta$.} But the combination $\overline{W} =W+\eta$ results
  a vanishing commutation relation  $
    [ \overline{W}({\bf x}), \overline{W} ({\bf y}) ] =0$ for all
  field points ${\bf x}$ and  ${\bf y}$ \cite{Dun:89}
  and hence $\overline{W}$ can be a diagonalized base for a representation
of the Schr\"odinger picture but not $W$ alone: Hence, the correct
representation is
\begin{eqnarray*}
  B({\bf y}) \left|\Psi \right> \rightarrow \frac{\hbar}{i \kappa}
  \frac{\delta}{\delta \overline{ W} ({\bf y})} \Psi (\overline{W})
\end{eqnarray*}
instead of (\ref{eq:naive_rep}). In this representation, it is easy to
see that the 
physical wavefunctional $\Psi_{\mbox{phys}}$ (\ref{eq:Gauss_law}),
which satisfies the Gauss' 
law constraint, is any functional made of $\overline{\phi}\equiv \phi e^{i
  \overline{W}}$ and $\overline{\phi}^{\dagger}\equiv \phi^{\dagger} e^{-i
  \overline{W}}~ [: J_0 : \equiv i \hbar (\phi \pi -\phi^{\dagger} 
\pi^{\dagger})]$\footnote{In Ref.\cite{S.K.Kim:90}, the forbidden
  combinations also were introduced explicitly and used to analyze the
  gauge equivalence.}  
\begin{eqnarray*}
  \left[\kappa B({\bf x})-:J_0 ({\bf x}): \right] \Psi_{\mbox{phys}}& =&
\left[ \frac{\hbar}{i} \frac{\delta}{\delta \overline{W}({\bf x})} -
  \hbar\phi 
    ({\bf x}) \frac{\delta}{\delta \phi ({\bf x})} +\hbar\phi^{\dagger}
    ({\bf x}) \frac{\delta}{\delta \phi^{\dagger} ({\bf x})} \right]
  \Psi_{\mbox{phys}}  \\
   &=&0 \nonumber
\end{eqnarray*}
with the usual representation for the matter parts
\begin{eqnarray}
\label{eq:matter_rep}
  &&\pi ({\bf x}) \left| \Psi_{\mbox{phys}} \right> \rightarrow
  \frac{\hbar}{i} 
  \frac{\delta}{\delta \phi ({\bf x})} \Psi (\phi, \phi^{\dagger} ) ,
  \nonumber \\
&&\phi ({\bf x}) \left| \Psi_{\mbox{phys}} \right> \rightarrow \phi
  ({\bf x}) \Psi (\phi, \phi^{\dagger} ) , 
  \nonumber \\
&&\pi^{\dagger} ({\bf x}) \left| \Psi_{\mbox{phys}} \right>
  \rightarrow \frac{\hbar}{i} 
  \frac{\delta}{\delta \phi^{\dagger} ({\bf x})} \Psi (\phi,
  \phi^{\dagger} ) , 
  \nonumber \\
&&\phi^{\dagger} ({\bf x}) \left| \Psi_{\mbox{phys}} \right>
  \rightarrow \phi^{\dagger} ({\bf x}) \Psi (\phi, \phi^{\dagger} ) 
\end{eqnarray}
; when the zero-mode is neglizable, $\overline{\phi},
\overline{\phi}^{\dagger}$ are nothing but the usual ones in
(\ref{eq:phys_variable_usual}) from 
the relation (\ref{eq:eta+W}). But our setting is more general than the
usual one: In particular, in a polynomial representation, it reads
\begin{eqnarray*}
  \Psi_{\mbox{phys}} ( \overline{\phi}, \overline{\phi}^{\dagger} ) = 
\prod_{j}^b
  \overline{\phi}({\bf x}_j) \prod_{i}^a \overline{\phi}^{\dagger} ({\bf
    y}_i) 
\end{eqnarray*}
and there is no factor of the purely gauge field, in this case
$\epsilon_{ij} \partial^{-1}_j B$ instead of ${\cal A}_i$. This
corresponds to a different representation from (\ref{eq:phys_wave_1})
and so this 
wavefunctional is not equivalent to (\ref{eq:phys_wave_1}) in general:
Actually, there 
is a slight difference in the purely gauge field part compared to
(\ref{eq:phys_wave_1}). For two particles sector, for example it becomes 
\begin{eqnarray}
  \label{eq:phys_wave_2}
  \Psi_{\mbox{phys}} &=&\overline{\phi}({\bf x}) \overline{\phi}({\bf
  y}) \nonumber
  \\
 &=&e^{i\eta ({\bf x})} e^{i\eta ({\bf y})} e^{[W({\bf x}), \eta({\bf
  y}) ] + \frac{1}{2} [ \eta({\bf x}), W ({\bf x})]+ \frac{1}{2} 
 [ \eta({\bf y}), W ({\bf y})] } \hat{\phi} ({\bf x})\hat{\phi} ({\bf y}).
\end{eqnarray}
by separating $e^{i \eta}$ and $\hat{\phi}$ parts: In the
  functional Schr\"odinger approach, the full ${\cal A}^k_{op}$ part, which
  has the momenta $B$ as well as $\eta$, is  not allowed in the form of
  (\ref{eq:phys_wave_1}) but only $\eta$ part is
  allowed and contributes to the physical states in the representation
of (\ref{eq:phys_wave_1}), in a specific from as
  (\ref{eq:phys_wave_2}). Finally, we note that the time-independent
  functional 
  Schr\"odinger equation becomes
  \begin{eqnarray*}
    \label{eq:Scheq}
   && H \Psi_E (\overline{\phi},\overline{\phi}^{\dagger})\nonumber \\
   &=&\int d^2 {\bf x}~
    : \left[ \overline{\pi}~\overline{\pi}^{\dagger} +
\overline{D^i \phi}(\overline{D^i \phi})^{\dagger} +m^2
    \overline{\phi} ~
\overline{\phi}^{\dagger} \right] : \Psi_E
   (\overline{\phi},\overline{\phi}^{\dagger}) \nonumber \\ 
  &=& \int d^2 {\bf x}~ \left[ -\hbar^2 \frac{\delta}{\delta
    \overline{\phi} ({\bf x})}\frac{\delta}{\delta
    \overline{\phi} ({\bf x})^{\dagger}} +\partial^i \overline{ \phi}
\partial^i \overline{ \phi}^{\dagger}
+ i \left( \overline{\phi}
    \partial^i \overline{\phi}^{\dagger} - \partial^i
    \overline{\phi}~\overline{\phi}^{\dagger} \right) {A^i}^{tr} + 
\overline{\phi}~\overline{\phi}^{\dagger}({A^i}^{tr})^2  \right]
    \Psi_E ( \overline{\phi}, \overline{\phi}^{\dagger}) \nonumber \\
&=& E \Psi_E ( \overline{\phi}, \overline{\phi}^{\dagger}),
  \end{eqnarray*}
where the barred variables in the Hamiltonian is the quantities
involved $\overline{\phi}, \overline{\phi}^{\dagger}$ and 
$
{A^i}^{tr} \equiv \epsilon_{ij} \partial_j^{-1} B 
=\frac{\hbar}{i \kappa} \epsilon_{ij} \partial_j^{-1}
\frac{\delta}{\delta \overline{W} ({\bf x})}  
$
which generates $(-) \frac{\hbar}{\kappa} \epsilon_{ij}
\partial_j^{-1} \delta^2 ({\bf x}-{\bf y}) =(-)\frac{\hbar}{2 \pi
  \kappa} \epsilon_{ij} \frac{({\bf x}-{\bf y})^j}{ |{\bf x}-{\bf
    y}|^2 } $ for each $\overline{ \phi} ({\bf y}) (\overline
{ \phi}^{\dagger} ({\bf y}))$ in the physical wavefunctional $\Psi_{E}
( \overline{\phi}, \overline{\phi}^{\dagger})$; this reflects the
relation ${A^i}^{tr} \approx
\frac{1}{\kappa}\epsilon_{ij} \partial^{-1}_j :J_0:$.
\begin{center} 
  {\bf D. Poincar\'e algebra}
\end{center}

As an important criterion of the consistency of the model, let us
consider the Poincar\'e algebra. In general, the quantum algebra
can have some anomaly term compared to the classical one
\cite{Park:94}
. But it is
found that this is not the case in our model: Classical algebra and
quantum algebra are the same with appropriate choice of ordering and
prescription. To see this, we first note, after some calculation, that
one can obtain the relation which is the most non-trivial one in
the Dirac-Schwinger conditions as follows, 
\begin{eqnarray*}
  \left[ T^{00}_s ({\bf x}),T^{00}_s ({\bf y}) \right]
  =\left( T^{0i}_s ({\bf x})+ T^{0i}_s ({\bf y}) \right) \partial^i_x
  \delta^2 ({\bf x}-{\bf y}).
\end{eqnarray*}
Using this condition, it is straightforward to find the following quantum
Poincar\'e algebra
\begin{eqnarray}
\label{eq:Poincare_alg}
&&[ P^{\mu}_s, P^{\nu}_s ] =0 , \nonumber \\
&&[ P^{\mu}_s, M^{\kappa \lambda}_s ] =i \hbar (\eta ^{\mu \lambda}
P^{\kappa} - \eta^{\mu \kappa} P^{\lambda} ) , \nonumber \\
&&[M^{\mu \nu}_s, M^{\kappa \lambda}_s ] = i \hbar \left(\eta^{\mu \kappa}
M^{\nu \lambda}- \eta^{\nu \kappa} M^{\mu \lambda}+ \eta^{\nu \lambda} 
M^{\mu \kappa}- \eta^{\mu \lambda} M^{\nu \kappa} \right)
\end{eqnarray}
as well as the classical one. Here, we have
considered symmetric ordering for the quantum generators in
(\ref{eq:Poincare_alg}) and 
used the condition of finite matrix elements of the Poincar\'e generators. 

On the other hand, there are canonical (Noether) Poincar\'e generators
which are usually identical to the improved ones
(\ref{eq:Poincare_gen}) {\it on the constraint surface} when one 
drops the boundary terms. But this is not trivial matter in lower
dimensions like as in (2+1)-dimensions since in that case the boundary
term should be treated more carefully \cite{Park:98b}
. Moreover, in the
Chern-Simons theory the situation is more serious: The gauge varying
``bulk'' terms also appear in the canonical
generators contrast to the improved ones. Let us describe this in
detail. The (classical) canonical 
Poincar\'e generators are defined as 
\begin{eqnarray*}
  &&P^{\mu}_c =\int d^2 {\bf x} ~ T^{0 \mu} _c , \nonumber \\
  &&M^{\mu \nu}_c =\int d^2 {\bf x} ~ \left[ x^{\mu} T^{0 \nu} _c-
  x^{\nu} T^{0 \mu} _c 
 + \pi_{\alpha} \Sigma ^{\mu \nu}_{\alpha \beta} A^{\beta} \right]
\end{eqnarray*}
with the canonical energy-momentum tensor
\begin{eqnarray*}
T^{\mu \nu}_{c}&=& \sum _{F=\phi, \phi^{*}, A_{\mu}} \frac{\partial
  {\cal L}}{\partial (\partial_{\mu}F)} \partial^{\nu}F -\eta^{\mu
  \nu} {\cal L} \nonumber \\
&=&(D^{\mu} \phi)^{*}(D^{\nu} \phi)+(D^{\mu} \phi)(D^{\nu}
  \phi)^{*}-J^{\mu}A^{\nu}-\eta^{\mu \nu}\left[(D^{\sigma}
  \phi)^{*}(D_{\sigma} \phi)-m^{2}\phi^{*} \phi \right]  \nonumber \\ 
&&~~~~+\frac{\kappa}{2} \eta^{\mu \delta} \epsilon_{\sigma \delta
  \rho} A^{\sigma} \partial^{\nu} A^{\rho} -\frac{\kappa}{2}\eta^{\mu
  \nu}\epsilon_{\sigma \eta \rho}A^{\sigma} \partial^{\eta} A^{\rho} 
\end{eqnarray*}
in the covariant form [$\pi_{\alpha} \equiv (\pi_0,\pi_i)=(0,
\frac{\kappa}{2} \epsilon_{ij} A^j )$] or
\begin{eqnarray*}
&&T^{00}_{c}=| \pi_{\phi} |^{2} +|D^{j} \phi|^{2}+m^{2}
|\phi|^{2}-A^{0}(J^{0}-\kappa B), \nonumber \\ 
&&T^{0i}_{c}=\pi_{\phi}D^{i}\phi +(D^{i} \phi)^{*} \pi_{\phi}^{*}
-J^{0}A^{i}- \frac{\kappa}{2}\epsilon_{jk}A^{j} \partial^{i} A^{k}
\end{eqnarray*}
in the components form. More explicitly, the canonical Poincar\'e
generators become
\begin{eqnarray*}
&&{P}^{0}_{c} = \int d^{2}{\bf x} \left[ |\pi_{\phi}|^{2} +|{D}^{i} \phi|^{2}+m^{2}|\phi|^{2}- A^{0}(J^{0}-\kappa B) \right], \nonumber \\
&&{P}^{i}_{c} =\int d^{2} {\bf x} \left[\pi_{\phi}{D}^{i}\phi
  +({D}^{i} \phi)^{*} \pi_{\phi}^{*} -J^{0}A^{i}-
  \frac{\kappa}{2}\epsilon_{jk}A^{j} \partial^{i} A^{k}\right],
\nonumber \\ 
&&{M}^{12}_{c} =  \int d^{2}{\bf x} \left
  [ \epsilon_{ij}x^{i}T^{0j}_{c}+ \epsilon_{ij} \pi^{i} A^{j} \right],
\nonumber \\ 
&&{M}^{0i}_{c} = x^{0} {P}^{i}_{c}-
 \int d^{2}{\bf x} \left[x^{i}T^{00}_{c}-\pi^{0}A^{i}+\pi^{i} A^{0} \right].
\end{eqnarray*}
Then, one can easily find the following relations between the
canonical and improved generators
\begin{eqnarray}
\label{eq:can_vs_imp}
&&{P}^{0}_{c} \approx {P}^{0}_{s}, \nonumber \\
&&{P}^{i}_{c} \approx {P}^{i}_{s}, \nonumber \\
&&{M}^{12}_{c} \approx {M}^{12}_{s} +\frac{\kappa}{2}\int d^{2} {\bf x}~
\partial^{k} \left(x^{k} {A}^{l} {A}^{l}- x^{l} {A}^{k} {A}^{l}\right)
, \nonumber \\ 
&&{M}^{0i}_{c} \approx {M}^{0i}_{s} +\frac{\kappa}{2} \int d^{2} {\bf
  x}~ A_{0} \epsilon^{ij} 
A^j ,
\end{eqnarray}
where we have dropped the boundary term
$
  \int d^2 {\bf x}~ \partial^j \left(\epsilon_{jk} A^i A^k \right),
$
which vanishes for the finite gauge field part of $P_c ^i$ or $M^{12}_c$;
however, the boundary term in `$M^{12}_c -M^{12}_s$' can not be simply
neglected. Here, one can observe that the canonical boost generator
$M^{0i}_c$ is not gauge invariant because of term $\frac{\kappa}{2}
\int d^{2} {\bf x}~ A_{0} \epsilon^{ij} A^j $ although the term `$M^{12}_c
-M^{12}_s$' is gauge invariant for the rapidly decreasing gauge
transformation function $\Lambda$ asymptotically. Moreover, the
commutators involving 
$M^{0i}_c$ do not satisfy the Poincar\'e algebra:
\begin{eqnarray*}
\frac{1}{i \hbar} [{M}^{0i}_{c}, {P}^{0}_{c}] &\approx&
  -{P}^{j}_{c}+\frac{\kappa}{2} \epsilon_{ik} \int d^{2}{\bf x}~ \partial^{0}
  \left(A^{0} {A}^{k}\right), \nonumber \\
\frac{1}{i \hbar}[{M}^{0i}_{c}, {P}^{j}_{c}] &\approx&
   -\delta_{ij} {P}^{0}_{c}+\frac{\kappa}{2} \epsilon_{ik} \int d^{2}
   {\bf x} ~\partial^{j} \left(A^{0} {A}^{k}\right), \nonumber \\
\frac{1}{i \hbar}[{M}^{0i}_{c}, {M}^{12}_{c}] &\approx&
 -\epsilon_{ij} {M} ^{0j}_{c}-\frac{\kappa}{2} \epsilon_{ij} x^0 \int
 d^{2} {\bf x}~ \partial^{l} \left(\epsilon_{lk}A_{j} {A}^{k}\right),
 \nonumber \\ 
\frac{1}{i \hbar}[{M}^{0i}_{c}, {M}^{0j}_{c} ] &\approx&
  -\epsilon_{ij} {M}^{12}_{c}-\frac{\kappa}{2} \epsilon_{ij} \int
  d^{2} {\bf x}~ \partial^{k} \left(x^{k}{A}^{l} {A}^{l}-x^{l}{A}^{k}
  {A}^{l}\right)\nonumber \\
 && -\frac{\kappa}{2} \epsilon_{ij} \int d^{2}{\bf x}~  
 \left[ \frac{5}{2} A^{2}_{0} +{A}^{k} {A}^{k} +\partial^{0}(x^{k}
   A^{0}{A}^{k}) \right].
\end{eqnarray*}
Here, we have also used the symmetric ordering for the quantum Poincar\'e
generators and used the prescription $\partial_i \delta^2 (0) \equiv 0$
in order to remove the undesirable infinities which arise from the
non-commuting $A_i$ at the same points; with these choices the quantum
algebras are the same as the classical ones. 

In conclusion, it is the improved
generators (\ref{eq:Poincare_gen}), constructed from the symmetric
energy-momentum tensor, which are (manifestly) gauge invariant and
obey the quantum as well as classical Poincar\'e algebra.  Hence these
improved generators have a unique meaning consistently with
Einstein's theory of gravity\footnote{There may be other differently
  improved generators 
  depending on what gravity theory is chosen like as in
  Ref. \cite{Cal:70}. But
  we do not consider this possibility in this paper. Moreover, the
  preferred property of  
  the symmetric energy-momentum tensor compared to
canonical one by the gauge invariance was examined also by Deser and
MaCarthy \cite{Des:90}. But they missed the important 
role of the gauge invariance, which is genuine for the CS gauge theory, 
on the integrated quantities, Poincar\'e generators.}; this will lead
to the uniqueness of 
the anomalous spin of the relativistic matter, which comes only from
$M^{12}_s$. (Detailed discussion will be presented in Sec. IV.) 

\begin{center}
{\bf IV. Matching of GIF and GFF} 
\end{center}

So far, we have considered the manifestly gauge invariant formulation
by introducing the Dirac dressing function. Now, the interesting
question is how the gauge invariant results are matched to gauge
fixed results. Actually, there have been some confusions about this
issue \cite{Dir:55,Lav:93,Gae:97}. This will be clarified in the
subsection {\bf D} and now we 
start by describing the correct matching to GFF which has been
presented recently \cite{Park:98a}.

In order to perform the matching, we need two things. One is the 
formula, called master formula 
\begin{eqnarray}
\label{eq:master_eq}
\{ L_a,~ L_b \} \approx \{ L_a ,~ L_{b} \} _{D_{\Gamma}}.
\end{eqnarray}
[The proof is presented in the Appendix {\bf C} and only the
interpretation of the result is in order here.]
Here $L_a$ is any gauge invariant quantity, where the bracket with the
first-class constraint $T$ of (\ref{eq:T}) vanishes, $ \{L_a,T\} \approx 0$.
The left-hand side of the formula (\ref{eq:master_eq}) is the basic
bracket of $L_a$'s. 
The right-hand side of (\ref{eq:master_eq}) is the Dirac bracket with
gauge fixing function 
$
\Gamma=0, ~~det |\{\Gamma, T\} | \neq 0.  
$
Moreover, in the
latter case, since $\Gamma=0$ can be strongly implemented, $L_a$ can
be replaced by ${L_a}|_{\Gamma}$ that represents the projection of
$L_a$ onto the surface ${\Gamma}=0$. 
The left-hand side is gauge independent by construction
since $L_a$'s and the the basic bracket algebra
(\ref{eq:basic_bracket}) are introduced 
gauge independently.  On the other hand, the Dirac bracket [9] depends
explicitly on the chosen gauge $\Gamma$ in general. But there is one
exceptional case, i.e., when the Dirac bracket is considered for the
gauge invariant variables. Our master formula (\ref{eq:master_eq})
explicitly show 
this exceptional case: The Dirac bracket for the gauge invariant
variables $L_a$ or their projection ${L_a}|_{\Gamma}$ on the surface
$\Gamma=0$ are still gauge invariant and equal to the basic bracket
for the corresponding variables\footnote{ For the first-class
  constraint $T$, it was known in Ref. \cite{Hen:92}; this formula was
implicitly included also
in the recently developed Batalin-Fradkin-Tyutin formalism
\cite{Bat:91}. But, the
  formula (\ref{eq:master_eq}) is valid even for the second-class
  constraint $T$ and 
  this fact has done an important role in the Dirac's canonical
  analysis of the boundary CS theory consistently to the
  symplectic reduction method \cite{Oh:98}.}. 

Another important thing for the
matching is to know how the defining equation (\ref{eq:dressing}) for
$c_{k}({\bf 
  x}-{\bf y})$ is modified in GFF. By considering $\hat{\phi}$ (or
${\cal A}_{\mu} )$ in a specific gauge and the residual gauge
transformation of $\phi$ and 
$A_{\mu}$, one can find modified (but still make $\hat{\phi}$ be gauge
invariant) equation for $c_k({\bf x}-{\bf y})$. Here we will consider
the following three
typical cases with the resultant modified equations of $c_k({\bf
  x}-{\bf y})$: 
\begin{eqnarray}
\label{eq:GFF}
&&\mbox{a)~Coulomb~ gauge}~(\partial^i A_i \approx 0)~:~ \int d^2 {\bf
  z}~ c_{j}
  ({\bf x}-{\bf z}) A^j({\bf z}) =0,\nonumber \\
&&\mbox{b)~Axial ~gauge}~(A_1 \approx 0)~:~ \partial^2_z c_2({\bf x}-{\bf z})
  =-\delta^2 ({\bf x}-{\bf z}),\nonumber \\
&&\mbox{c)~Weyl ~gauge}~(A_0 \approx 0)~:~ \partial^j_z c_j({\bf
  x}-{\bf z})= -\delta^2 ({\bf x}-{\bf z}).
\end{eqnarray}
These results are generally valid for any other gauge theories when
they are formulated by our gauge invariant formulation.  Note that
these results are different from recent claims of
Ref. \cite{Lav:93} except in the case of Coulomb gauge\footnote{ Authors of
Refs. \cite{Lav:93,Gae:97} considered $\int d^2 {\bf z}~ c_j ({\bf
    x}-{\bf z} ) A^j ({\bf z}) =0$ even ``b)'' and ``c)'' cases. But
  then, the manifestly gauge invariant fields in
  (\ref{eq:phys_variable}) are not gauge 
  invariant under the residual gauge symmetries $ \phi \rightarrow
  e^{-i\Lambda } \phi,~A_{\mu} \rightarrow A_{\mu} +\partial_{\mu}
  \Lambda$ with $x^1$ and $x^0$ independent $\Lambda$ for ``b)'' and
  ``c)'', respectively.}.  Moreover, the Weyl gauge does not
modify the equation for $c_k$ from (\ref{eq:dressing}).  Then, using
these relations 
and (\ref{eq:master_eq}), we can consider the gauge fixed results
directly from GIF. 
However, as can be observed in these examples, gauge fixings restrict
the solution space in general. Therefore, all the variables which
appear in (\ref{eq:ano_transf}) are gauge invariant for each solution
hyper-surface which 
is selected by gauge fixing, but their functional form may be
different depending on the chosen gauges. Let us derive the results of
(\ref{eq:GFF}) 
in detail and the matching to GFF using them.

\begin{center}
  {\bf A. Coulomb gauge $(\partial^i A_i \approx 0)$ }
\end{center}

It is important to note that in this gauge, there is no residual gauge
symmetry: If we consider the gauge transformation, i.e., $A_i
\rightarrow A_i +\partial_i \Lambda$, $\Lambda$ should satisfy the
Laplace equation
$
  \nabla^2 \Lambda ({\bf x}) =0
$
over all space-times and so this equation has only one trivial
solution
$
  \Lambda ({\bf x})=0.
$
On the other hand, since the gauge transformation of $\phi$ is defined
as what makes $D_i
=\partial_i + i A_i$ as a covariant derivative when acted upon $\phi$,
$\phi$ does not transform (modular unimportant global
phase transformation), either. Hence, it is found that the
additional factor $W$, which cancels the gauge transformations of
$A_i$ and $\phi$, is unnecessary or it can be made to be zero simply,
i.e.,
\begin{eqnarray}
\label{eq:c_Coulomb}
W( {\bf x})=\int d^{2} {\bf z}~ c_{j}({\bf x}, {\bf z})A^{j}({\bf z})=0 . 
\end{eqnarray}
In this case, the gauge invariant variables in (\ref{eq:phys_variable})
and the corresponding base
fields are equivalent, i.e., ${\cal A}_{\mu} =A_{\mu}, \hat{\phi} =\phi,
  \hat{\pi}=\pi$ and thus the solution of $A_{\mu}$ in this gauge can
  be directly read from (\ref{eq:AA_i_solution}), (\ref{eq:AA_0_solution}) as
  follows \cite{Hag:84}\footnote{Here, one should be careful in order
    not to obtain a wrong result by applying ${\cal 
      A}_{\mu} \rightarrow A_{\mu}$ before applying the constraint or
    equations of motion: Only for the final formula in
    (\ref{eq:AA_i_solution}) and (\ref{eq:AA_0_solution}), one can get
    the correct results.}:
    \begin{eqnarray*}
      &&A_i ({\bf x}) \approx -\frac{1}{2 \pi \kappa} \int d^2 {\bf z}~
      \epsilon_{ik} \frac{({\bf x}- {\bf z})_k}{|{\bf x}- {\bf z}|^2}
      J^0 ({\bf z}), \nonumber \\
    &&A_0 ({\bf x}) = -\frac{1}{2 \pi \kappa} \int d^2 {\bf z}~
      \epsilon_{kj} \frac{({\bf x}- {\bf z})_k}{|{\bf x}- {\bf z}|^2}
      J^j ({\bf z}).
    \end{eqnarray*}

Now, in order to find the solution $c_j$ of (\ref{eq:c_Coulomb}), let
us define $c_j$ 
as
\begin{eqnarray}
\label{eq:c_special} 
 c_j ({\bf x}-{\bf z}) =\partial_j ^z \chi ({\bf x} -{\bf z})
\end{eqnarray}
then, it is easy to see
\begin{eqnarray*}
\int d^{2} {\bf z}~ c_{j}({\bf x}-{\bf z})A^{j}({\bf z}) 
&&=\int d^{2} {\bf z}~ \partial_j^z \chi({\bf x}-{\bf z})A^{j}({\bf
  z}) \nonumber \\ 
&&=-\int d^{2} {\bf z}~  \chi({\bf x}-{\bf z})\partial_j^zA^{j}({\bf
  z}) \nonumber \\
&&=0
\end{eqnarray*}
when one neglects the boundary term, i.e., 
\begin{eqnarray}
\label{eq:boundary}
\int d^{2} {\bf z}~ \partial_j^z \left(\chi({\bf x}-{\bf z})A^{j}({\bf
  z})\right)=0.  
\end{eqnarray}
Now, to find the solution of $\chi$ with this property, let us note
that $\chi$ satisfies the Poisson equation
$
  \nabla^2 \chi({\bf x}-{\bf z}) = \delta^2 ({\bf x}-{\bf z}) 
$
according to (\ref{eq:dressing}). The well-known solution of this equation is
\begin{eqnarray}
\label{eq:Green_fun}
\chi({\bf x}-{\bf z}) =\frac{1}{4 \pi} ln |{\bf x}-{\bf z}|^2 
\end{eqnarray}
up to unimportant constant term. It is easy to find that this solution
satisfies (\ref{eq:boundary}) as 
\begin{eqnarray*}
\int d^{2} {\bf z}~ \partial_j^z \left(\chi({\bf x}-{\bf z})A^{j}({\bf
  z})\right)  
&&=\oint_{S^1_{R \rightarrow \infty}} R d \theta \hat{r} \cdot
  \left(\chi({\bf 
  x}-{\bf z})A^{j}({\bf z})\right) \nonumber \\
&&=\frac{Q}{4 \pi^2 \kappa} \oint_{S^1_{R \rightarrow \infty}} d
  \theta \hat{ r} 
  \cdot \hat{ \theta}  ln R \nonumber \\
&&=0  
\end{eqnarray*}
``geometrically'' though not neglizable in the naive asymptotic
$r$-dependence. [Here the 
integration is evaluated on a circle with infinite radius $R$, polar
angle $\theta$, and their corresponding (orthogonal) unit vectors
$\hat{r},~\hat{\theta}$.] Hence, one finds the solution, by using
(\ref{eq:c_special}) and (\ref{eq:Green_fun}), as follows
\begin{eqnarray}
\label{eq:c_solution_Coul}
c_j({\bf x}-{\bf z}) =-\frac{1}{2 \pi}
 \frac{({\bf x}-{\bf z})_j}{|{\bf x}-{\bf z}|^2}.
\end{eqnarray}
Then, we find the anomalous
transformation (\ref{eq:ano_transf}) with
\begin{eqnarray}
\label{eq:ano_factor_Coulomb}
\Xi^{12}({\bf x})&=&\frac{1}{2 \pi \kappa} Q, \nonumber \\
 \Xi^{0i}({\bf x})&=&
\frac{1}{2 \pi \kappa} \int d^2 {\bf z}~ \frac{({\bf x} -{\bf z})^i
 ({\bf x}-{\bf z})^k}{|{\bf x}-{\bf z}|^2} \epsilon_{jk} J^j({\bf z}),
\end{eqnarray}
where $Q=\int d^2 {\bf z} J_0$, which can be directly obtained from the
expressions of (\ref{eq:ano_factor}) by substituting
(\ref{eq:c_solution_Coul}). These are exactly Hagen's rotational 
anomaly and Coulomb gauge restoring term in the Lorentz
transformation, respectively \cite{Hag:84}.  

Next, let us consider the two functions $\Delta ({\bf x}-{\bf y})$ and
$\xi_{ij} ({\bf x}-{\bf y})$ which characterize the commutation relations.
About $\Delta ({\bf x}-{\bf y})$ (\ref{eq:Delta}), one finds that,
upon using (\ref{eq:c_special}), 
\begin{eqnarray}
\label{eq:Delta=0} 
 \Delta({\bf x}-{\bf y})&=& \int d^2 {\bf z}~ \epsilon^{kj} \partial_k^z
  \chi ({\bf x}-{\bf y})\partial_j^z \chi ({\bf y}-{\bf z}) \nonumber
  \\
 &=& -\int d^2 {\bf z}~  \epsilon^{kj} \nabla^z \times \left[ \nabla^z
  \chi ({\bf x}-{\bf y}) \chi ({\bf y}-{\bf z}) \right] \nonumber
  \\
&=&-\oint_{S^1_{R \rightarrow \infty}} R d \theta \hat{\theta} \cdot
  \left[\nabla^r 
  \chi ({\bf x}-{\bf r}) \chi ({\bf y}-{\bf r}) \right] \nonumber
  \\
&=&-\frac{1}{2 \pi} \oint_{S^1_{R \rightarrow \infty}}  d \theta \hat{\theta}
  \cdot \hat{r} ln R \nonumber \\
&=&0
\end{eqnarray}
from the geometric reason, by performing the integration by parts in
  the second line.  Now, about $\xi_{ij}({\bf x}-{\bf y})$ (\ref{eq:Xi}), one
  finds that, upon using the antisymmetry $c_j({\bf
  x}-{\bf z})=-c_j({\bf z}-{\bf x})$ for the solution
  (\ref{eq:c_solution_Coul})  
\begin{eqnarray*}
  \xi_{ij}({\bf x}-{\bf y})&=&-\epsilon_{ik} \partial_j^y c_k({\bf
  y}-{\bf x})
+\epsilon_{kj} \partial_i^x c_k ({\bf x}-{\bf y}) \nonumber \\
 &=&\left(\epsilon_{ik} \partial_j^x  
+\epsilon_{kj} \partial_i^x \right) c_k ({\bf x}-{\bf y})
\end{eqnarray*}
which becomes, for each indices, as follows:
\begin{eqnarray*}
  a)\left(\epsilon_{ik} \partial_i^x  
+\epsilon_{ki} \partial_i^x \right) c_k ({\bf x}-{\bf y}) &=&0
~~~(\mbox{for}~ i=j), \nonumber \\
  b)\left(\epsilon_{1k} \partial_2^x  
+\epsilon_{k2} \partial_1^x \right) c_k ({\bf x}-{\bf y})
&=&\partial_2^x  c_2 ({\bf x}-{\bf y})+\partial_1^x  c_1 ({\bf x}-{\bf
  y})\nonumber \\
&=&-\delta^2 ({\bf x}-{\bf y})~~~ (\mbox{for} ~i=1, j=2), \nonumber \\
  c)\left(\epsilon_{2k} \partial_1^x  
+\epsilon_{k1} \partial_2^x \right) c_k ({\bf x}-{\bf y}) 
&=&-\partial_1^x  c_1 ({\bf x}-{\bf y})-\partial_2^x  c_2 ({\bf x}-{\bf
  y})\nonumber \\
&=&\delta^2 ({\bf x}-{\bf y})~~~ (\mbox{for}~ i=2, j=1). \nonumber \\
\end{eqnarray*}
In a compact form, it becomes 
\begin{eqnarray*}
\xi_{ij} ({\bf x}-{\bf y})=-\epsilon_{ij} \delta^2({\bf x}-{\bf y}).  
\end{eqnarray*} 
Using these results, one finds that the basic brackets
defined in (\ref{eq:graded_bracket}) are the usual Dirac brackets in
the Coulomb gauge 
\begin{eqnarray*}
\{ {\cal A}_i({\bf x}),{\cal A}_j({\bf y}) \}_{D(Coulomb)} &\approx&
\{ {A}_i({\bf x}),{A}_j({\bf y}) \}_{D(Coulomb)}=0 , \nonumber \\
\{ {\hat{\phi}}({\bf x}),{\hat{\phi}}({\bf y}) \}_{D(Coulomb)} &\approx&
\{ {{\phi}}({\bf x}),{{\phi}}({\bf y}) \}_{D(Coulomb)} =0 , \nonumber \\
\{{\cal A}_i({\bf x}), \hat{\phi}({\bf y}) \}_{D(Coulomb)} &\approx&
\{{A}_i({\bf x}), {\phi}({\bf y}) \}_{D(Coulomb)}=
  -\frac{i}{2 \pi \kappa} \epsilon_{ik} \frac{({\bf x}- {\bf
      z})_k}{|{\bf x}- {\bf z}|^2}{{\phi}}({\bf y}). 
  \end{eqnarray*}
Furthermore, these two results imply that the gauge
invariant operator $\hat{\phi}_{op}$ satisfies the boson commutation
relation, 
$
[ \hat{\phi}_{op}({\bf x}),\hat{\phi}_{op} ({\bf y})]=0  
$
in this case instead of the generic graded commutation relations
(\ref{eq:graded_comm}).  Here, we note the special importance of the
Coulomb gauge 
in that the original fields $\phi,~ \phi ^*, ~A_{\mu}$ themselves are
already gauge invariant fields because of the result
(\ref{eq:c_Coulomb}) and hence they already have the full 
anomaly structures of (\ref{eq:ano_transf}). Furthermore, this gauge
is the simplest one 
to obtain the anomalous spin of the original matter field $\phi$ as
$\Xi^{12}$ of (\ref{eq:ano_factor_Coulomb}) since this does not have
other gauge restoring terms 
as in the rotationally non-symmetric gauge. This is made clear by
noting the relation of (\ref{eq:can_vs_imp}) 
\begin{eqnarray*}
M^{12}_s \approx M^{12}_c -\frac{\kappa}{2} 
\int d^2 {\bf z} ~ \partial ^k \left(z^k A^l
A^l-z^l A^l A^k \right),
\end{eqnarray*} 
where $M^{12}_c$ is the canonical angular momentum
\begin{eqnarray*}
M^{12}_c =\int d^2 {\bf z}~ \left[ \epsilon_{lk} z^l \left(\pi
    \partial^k \phi +  
(\partial^k \phi)^* \pi^* \right) -\kappa z^l A^l (\partial^k A^k)
    +\frac{\kappa}{2} 
 \partial^k (z^l A^l A^k)\right].
\end{eqnarray*}
The surface terms in $M^{12}_s -M^{12}_c$ and $M^{12}_c$, which are
gauge invariant for the rapidly decreasing gauge transformation
function $\Lambda$, give the gauge independent spin terms ``$({1}/{4
  \pi \kappa}) Q^2$" \cite{Hag:84,Hag:85,Shin:92} (unconventional) and
``$0$" (conventional) 
in $M^{12}_s$, respectively: Because of the gauge independence of the
surface terms, only the calculation in one simple gauge, e.g., Coulomb
gauge is 
sufficient\footnote{Explicit manipulations of the gauge
independence of the unconventional term have been established only for some
limited class of gauges \cite{Hag:84,Hag:85,Shin:92}. But these
results will be generalized to 
the case of general gauges due to the gauge invariance of the term.}
to get this general 
result and in that case one obtains explicitly the boundary integrals
as follows  
\begin{eqnarray*}
  \int d^2 {\bf z}~ \partial^k \left(z^k A^l A^l \right) 
&=&-\oint_{S^1_{R \rightarrow \infty}} R d \theta \hat{r} \cdot {\bf
  r} ~A^l A^l 
\nonumber \\
&=&-\frac{1}{4 \pi \kappa^2 }\oint_{S^1_{R \rightarrow \infty}} R^2 d
\theta ~ 
 \epsilon_{lk} \frac{r^k}{|{\bf r}|^2} Q ~ \epsilon_{lj}
 \frac{r^j}{|{\bf r}|^2} Q 
\nonumber \\
&=&-\frac{1}{\kappa^2} Q^2 ,\nonumber \\
\int d^2 {\bf z}~ \partial^k \left(z^l A^l A^k \right) 
&=&-\oint_{S^1_{R \rightarrow \infty}} R d \theta \hat{r} \cdot {\bf r} \cdot
{\bf A} ~z^l A^l \nonumber \\
&=&0.
\end{eqnarray*}
From which the anomalous spin $\frac{Q}{2 \pi \kappa}$ of
(\ref{eq:ano_factor_Coulomb}) for the 
base matter field, which
is defined as the surface term $M^{12}_s -M^{12}_c$ for the
base matter field is readily seen to follow for general gauges: The
commutation relation  
$
[ Q, \phi ({\bf x}) ]=\phi ({\bf x})   
$
which is a basic ingredient in the derivation, 
is gauge independent relation because it expresses the gauge independent fact
that $\phi$ carries the unit charge\footnote{This can be explicitly
  checked by 
  considering the general gauges $\int d^2 {\bf z}~
  K_{\mu}({\bf x}, {\bf z}) A^{\mu}({\bf z}) \approx 0$ with a kernel
  $K_{\mu}({\bf x},{\bf z})$ as will be discussed in subsection {\bf
    D} in a different context.}. On
the other hand, the second term of $M^{12}_c$, which vanishes only in
the Coulomb gauge\footnote{The other factor $z^l A^l$ will not be
  zero for all space region.}, gives the gauge restoring
contribution to the rotation transformation for the matter
field for the general gauges; actually, this is the case for the
rotation non-invariant gauge 
since the Coulomb gauge only is the ``rotation invariant
'' one\footnote{These gauge should be ``translation'' invariant also such
  that the each gauge is defined over all space.}. Finally, we note
that the anomalous spin, which comes only from 
  $M^{12}_s$, has a unique meaning because of the uniqueness of the
  improved generators in that they are gauge invariant on the
  constraints surface and obey the Poincar\'e algebra though this is
  not the case for the canonical ones. 

Furthermore, this uniqueness of anomalous spin is in contrast to the
anomalous statistics, which has only artificial meaning in this case
\cite{Hag:89}. 
This is because we can obtain in any field theories  
any
$arbitrary$ statistics by constructing gauge invariant exotic operators
of the form of Semenoff and its several variations \cite{Hag:84}. In
this sense 
the relativistic CS gauge theory does not respect the
{\it spin-statistics} relation \cite{Wil:82} in agreement with Hagen's
result \cite{Hag:84,Hag:89}.  Here, we would like to comment that the
situation of non-relativistic CS gauge 
theory is not better than the relativistic case. This is because even
though the anomalous statistics is uniquely defined by removing the
gauge field (in this case the gauge field is pure gauge due to point
nature of the sources in non-relativistic quantum field theory) the
anomalous spin has no unique
meaning \cite{Hag:84,Hag:85}: The anomalous spin can be removed by
redefining the 
angular momentum generator without distorting the Poincar\'e algebra.

\begin{center}
  {\bf B. Axial gauge $(A_1 \approx 0)$}
\end{center}

In this gauge, the residual gauge symmetry is 
$
A_i \rightarrow A_i
=\partial_i \Lambda,~ \partial_1 \Lambda=0
$
 which preserves the chosen gauge $A_1 \approx 0$. Under this
 transformation, the matter field $\hat{\phi}$ is
 gauge invariant when the dressing satisfies the equation
 \begin{eqnarray}
\label{eq:dressing_axi_2}
 \partial^2_z c_2({\bf x}-{\bf z})
  =-\delta^2 ({\bf x}-{\bf z}) , 
 \end{eqnarray}
and by comparing to the original equation (\ref{eq:dressing}) one obtains
furthermore another equation
\begin{eqnarray}
\label{eq:dressing_axi_1}
 \partial^1_z c_1({\bf x}-{\bf z})
  =0 .
 \end{eqnarray}
Now, let us consider the several solutions of the equations
(\ref{eq:dressing_axi_2}) and (\ref{eq:dressing_axi_1}). First the
simplest solution is \cite{Dir:55}
\begin{eqnarray}
\label{eq:c_solution_axi}
  c_1({\bf x}-{\bf z}) &=& 0 ,\nonumber \\
  c_2({\bf x}-{\bf z})&=&-\delta(x_1-z_1) \epsilon (x_2-z_2) 
   \end{eqnarray}
 with the step function $\epsilon(x)$
 \begin{eqnarray*} 
   \epsilon(x) =\left\{ \begin{array}{ll} 0 &(x < 0 ) \\
                1 &  (x > 0 )\end{array} \right. .
 \end{eqnarray*}
This gives for $W$ a line integral\footnote{Here, we are considering
  the line-integral representation of $W$ in the context of gauge
  fixing which is well-defined over all space, the crossings in the
  contour of the integral is not considered. Hence, in our case, there
  is no path-ordering even when we consider the quantum theory
  contrast to the case of \cite{Dun:89}.}
\begin{eqnarray*}
  W( {\bf x})= \int ^{x_2} _{-\infty} d z_2 ~A^2 (x_1, z_2).
\end{eqnarray*}
In this case, $\hat{\phi}_{op} ({\bf x})$ carries the vector potential
at the position ${\bf y}$,
$
  a^1 ({\bf y}) = -\frac{\hbar}{\kappa} \delta(x_1-y_1) \epsilon
  (x_2-y_2), 
  a^2 ({\bf y}) =0 
$, which is orthogonal to and non-vanishing only along the integration
path; this can be considered as a shrink of the space where the gauge
field lives for the Coulomb gauge into one-dimensional (straight)
lineal space.
Moreover, since this solution (\ref{eq:c_solution_axi}) corresponds to
a different solution hyper-surface to the Coulomb 
gauge and therefore, its related anomalous terms in
(\ref{eq:ano_transf}) have different 
functional form to (\ref{eq:ano_factor_Coulomb}) even though they are
gauge invariant on their own hyper-surfaces: 
\begin{eqnarray*}
 \Xi^{01} ({\bf x}) &=&0, \nonumber \\
 \Xi^{02} ({\bf x}) &=&\frac{1}{\kappa} \int
^{\infty}_{-\infty} d y_2 ~ J_1 (x_1, y_2), \nonumber \\
 \Xi^{12} ({\bf x}) &=&-\frac{1}{\kappa} \int ^{\infty}_{-\infty}
d y_2 ~ J_0 (x_1, y_2).
\end{eqnarray*}
Here, we have used the formula (\ref{eq:ano_factor}).

Furthermore, one finds that the two characteristic functions of
(\ref{eq:Delta}) and (\ref{eq:Xi}) become as follows 
\begin{eqnarray}
\label{eq:Xi_axi}
  &&\Delta({\bf x}-{\bf y}) =0 , \nonumber \\
  &&\xi_{11} ({\bf x}-{\bf y}) =-\partial_1 ^x \delta (x_1 -y_1),
  \nonumber \\
   &&\xi_{12} ({\bf x}-{\bf y}) =-\xi_{21} ({\bf x}-{\bf y})=\delta^2
  ({\bf x} -{\bf y}), \nonumber \\
  &&\xi_{22} ({\bf x}-{\bf y}) = 0,
\end{eqnarray}
and thus the basic brackets of (\ref{eq:graded_bracket}), which are
found to be the Dirac 
bracket in the axial gauge $A_1 \approx 0$, become as follows
\begin{eqnarray*}
&&\{ {\cal A}_1({\bf x}),{\cal A}_1({\bf y}) \}_{D(axial)} =
  -\frac{1}{\kappa} \partial_1^x \delta({x}_1-{y}_1), \nonumber \\
&&\{ {\hat{\phi}}({\bf x}),{\hat{\pi}}({\bf y}) \}_{D(axial)} =
  \delta^2 ({\bf x}-{\bf y}), \nonumber \\
&&\{{\cal A}_1({\bf x}), \hat{\phi}({\bf y}) \}_{D(axial)} =
  -\frac{i}{\kappa}{\hat{\phi}}({\bf y})\delta({y}_1-
  {x}_1) \epsilon({y}_2-{x}_2) , \nonumber \\
&&\{{\cal A}_1({\bf x}), \hat{\pi}({\bf y}) \}_{D(axial)} =
  \frac{i}{\kappa}{\hat{\phi}}({\bf y})\delta({y}_1-
  {x}_1) \epsilon({y}_2-{x}_2) , \nonumber \\
&&\mbox{others~vanish},
\end{eqnarray*}
where
\begin{eqnarray}
\label{eq:phys_field_axi}
{\cal A}_1 ({\bf x}) &=& -\partial_1^x \int d^2 {\bf z}~ c_2 ({\bf
  x}-{\bf z}) A^2 ({\bf z}) \nonumber \\
 &=& - \partial_1^x \int^{x_2}_{-\infty} d z_2 ~A^2 (x_1, z_2 ) ,
  \nonumber \\
{\cal A}_2 ({\bf x}) &=& A_2 ({\bf x}) -\partial_2^x \int d^2 {\bf z}~
  c_2 ({\bf x}-{\bf z}) A^2 ({\bf z}) \nonumber \\
 &=& 0, \nonumber \\ 
\hat{\phi} ({\bf x}) &=&\phi ({\bf x}) e^{i \int^{x_2}_{-\infty} d z_2 ~
  A^2 (x_1, z_2 ) } , \nonumber \\ 
\hat{\pi} ({\bf x}) &=&\pi ({\bf x}) e^{-i \int^{x_2}_{-\infty} d z_2 ~
  A^2 (x_1, z_2 ) }.
\end{eqnarray}
However, it is not straightforward to obtain the Dirac bracket for the
base fields $\phi$ and $A^i$
themselves . Moreover, for more general
solution
\begin{eqnarray}
\label{eq:c_axi_gen}
c_1({\bf x}-{\bf z}) &=&G (x_2 -z_2 ) ,\nonumber \\
c_2({\bf x}-{\bf z})&=&\delta(x_1-z_1) \epsilon (x_2-z_2) + F(x_1 -z_1)  
\end{eqnarray}
with  the functions $F$ and $G$ which have
the dependence only along the $1$ and $2$ directions,
respectively, and the gauge invariant variables
(\ref{eq:phys_field_axi}) are not changed except
$
  {\cal A}_1 ({\bf x}) = \int d^2 {\bf z} ~ [ \partial^z_1 F (x_1
  -z_1) ] A^2 ({\bf z})
$
but all other variables $\Xi^{\mu \nu}, \Delta, \xi_{ij}$ have the explicit
dependence of two functions $F$ and $G$:
\begin{eqnarray*}
\Xi^{12} ({\bf x}) &=& - x^2 \int d^2 {\bf z} ~ \left[\partial^z_1 F
  (x_1 -z_1) \right] A^2 ({\bf z}) -\frac{1}{\kappa} \int
^{\infty}_{-\infty}dz_2 ~ J_0 (x_1, z_2 ), \nonumber \\
\Xi^{01} ({\bf x}) &=& \frac{1}{\kappa}  \int d^2 {\bf z} ~({\bf x}
  -{\bf z})_1 
\left[ G(x_2 -z_2) J^2 ({\bf z})- F(x_1 -z_1) J^1 ({\bf z}) \right], 
\nonumber \\
\Xi^{02} ({\bf x}) &=& \frac{1}{\kappa}  \int d^2 {\bf z} ~({\bf x}
  -{\bf z})_2 
\left[ G(x_2 -z_2) J^2 ({\bf z})- F(x_1 -z_1) J^1 ({\bf z}) \right] +
\frac{1}{\kappa}\int^{\infty}_{-\infty}dz_2 ~ J_1 (x_1, z_2 ),  
\nonumber \\
 \Delta({\bf x}-{\bf y}) &=& \int^{y_2}_{-\infty} d z_2 ~ G(x_2 -z_2) -
\int^{x_2}_{-\infty} d z_2 ~ G(y_2 -z_2) \nonumber \\
&&~~+ \int d^2 {\bf z} ~
\left[G(x_2 -z_2) F(y_1 -z_1) -F(x_1 -z_1)G(y_2 -z_2) \right],
\nonumber \\
\xi_{11}({\bf x}-{\bf y})&=&-\partial_1^x \delta (x_1-y_1) -\partial_1^x
\left[ F(y_1 -x_1) +F(x_1 -y_1) \right], \nonumber \\
\xi_{12}({\bf x}-{\bf y})&=&-\xi_{21}({\bf x}-{\bf
  y})=\delta^2({\bf x}-{\bf y}).  
\end{eqnarray*}
It is interesting to note that $\xi_{ij}({\bf x}-{\bf y})$ is the same
as (\ref{eq:Xi_axi}) 
if $F(x_1 -y_1)$ is an anti-symmetric function, i.e., 
$F(x_1 -y_1)=-F(y_1 -x_1)$. Moreover, we note that $\Delta({\bf
  x}-{\bf y})$ is not zero in general and hence the commutation
relations of $\hat{\phi}$'s are not the bosonic ones and we have additional
contribution in the commutation relation involving ${\cal A}_i ({\bf
  x})$ according to (\ref{eq:graded_comm}).

\begin{center}
  {\bf C. Weyl gauge $(A_0 \approx 0)$}
\end{center}

In this case, the residual symmetry is the time-independent gauge
transformation 
$
  A_{\mu} \rightarrow A_{\mu} + \delta_{\mu i} \partial_i \Lambda,
$
with the time-independent gauge function $\Lambda$. However, because
 of the form of Dirac dressing as (\ref{eq:W}) which is spatially
 non-local but temporally local, the gauge 
 transformation of the $\hat{\phi}, {\cal A}_{\mu}$ fields
 (\ref{eq:phys_variable}) is formally the same as 
 that of gauge unfixed case and this gives the same equation for
 the dressing as (\ref{eq:dressing}).
 
\begin{center}
{\bf D. Clarification of previous confusions }
\end{center}

Up to now, there have been several confusions about the gauge
fixing (kernel-) function $K_{\mu}({\bf x}, {\bf z})$ in the gauge
condition
\begin{eqnarray}
\label{eq:gen_gauge}
\int d^2 {\bf z}~ K_{\mu}({\bf x}, {\bf z}) A^{\mu} ({\bf z}) =0  
\end{eqnarray}
and the Dirac dressing function $c_k ({\bf x}-{\bf z})$ which
satisfies (\ref{eq:dressing}) \cite{Dir:55,Lav:93,Gae:97}. The main
source of these confusions is the identity \cite{Park:97,Dir:55} 
\begin{eqnarray}
\label{eq:identity}
\int d^2 {\bf x}~ c_i({\bf y}-{\bf x}) {\cal A}^i ({\bf x}) =0,  
\end{eqnarray}
where the function $c_i ({\bf y}-{\bf x})$ rapidly decreases in the
asymptotic region so that a boundary integral
$
  \int d^2 {\bf x} ~\partial_x^i \left[ c_i ({\bf y}-{\bf x})   W({\bf
x}) \right]
$
can be neglected. The identity (\ref{eq:identity}) looks similar to 
(\ref{eq:gen_gauge}), but there
is an important difference. In (\ref{eq:gen_gauge}), $A^{\mu}$ denotes
the gauge varying field 
and hence (\ref{eq:gen_gauge}) restricts the gauge symmetry of
$A^{\mu}$, while ${\cal
  A}^i$ in (\ref{eq:identity}) is gauge invariant by definition and
hence (\ref{eq:identity}) does 
not restrict the gauge symmetry; the identity might be involved with
some other 
symmetries if there are. Actually, it is found that a new symmetry has
been introduced implicitly  with the introduction of $c^i ({\bf y}-{\bf x})$
\cite{Dir:55}\footnote{It
  is interesting to compare this symmetry to the BRST symmetry
  \cite{Bec:76}. These two symmetries seem to be an extremely
  different ones; since one (BRST symmetry) is about GFF and the other
  ($c_k$-symmetry) is about GIF. But 
  they are rather very close in that the fundamental variables
  (${\cal F}_{\alpha}$ in the former and $(A_{\mu}, \phi, \phi^* )$
  in the latter ) are not gauge transformed: In the former case, the
  variable ${\cal F}_{\alpha}$ are not gauge transformed by
  definition; in the latter case, the 
  variables $(A_{\mu}, \phi, \phi^* )$ are not allowed to be gauge
  transformed because of the introduction of a complete gauge fixing term.} 
\begin{eqnarray}
\label{eq:c_sym}
c_k ({\bf x}-{\bf z}) \rightarrow c_k ' ({\bf x}-{\bf z}) 
=c_k ({\bf x}-{\bf z})+
\epsilon_{kj} \partial ^z _j b ({\bf x}-{\bf z}) 
\end{eqnarray}
because the transformed quantity $c_k ' ({\bf x}-{\bf z}) $ also
satisfies the defining equation (\ref{eq:dressing}) for the
well-behaved function  
$b ({\bf x}-{\bf z}) $ with $\nabla \times \nabla  b 
=0$. However, this transformation is not trivial one because it
makes $ W$ transform as follows
\begin{eqnarray}
\label{eq:W_sym}
  W ({\bf x}) \rightarrow W ({\bf x}) + \int d^2 {\bf z} ~b ({\bf
  x}-{\bf z}) B({\bf z})
\end{eqnarray}
by neglecting the boundary term
$
  \int d^2 {\bf z} ~\partial ^z _j \left[ \epsilon_{kj} b({\bf x}-{\bf
  z}) A^k ({\bf z}) \right]
$
for rapidly decreasing function $b ({\bf x}-{\bf
  z})$. Hence one finds a strange situation that the gauge invariant
variables 
${\cal F}^{\alpha}$ transform as follows
\begin{eqnarray}
\label{eq:new_sym}
  &&\hat{\phi} ({\bf x}) \rightarrow \hat{\phi}' ({\bf x})=e^{i
  \Lambda({\bf x})} \hat{\phi} ({\bf x}), \nonumber \\
  && {\cal A}_{\mu}({\bf x}) \rightarrow {\cal A}_{\mu} '({\bf x})
  ={\cal A}_{\mu}({\bf x}) -\partial_{\mu}^x   \Lambda({\bf x}),
\end{eqnarray}
which looks like as a gauge transformation with a transformation
function 
$
\Lambda ({\bf x} ) = -  \int d^2 {\bf z} ~b ({\bf
  x}-{\bf z}) B({\bf z}) 
 \approx - \frac{1}{\kappa} \int d^2 {\bf z} ~b ({\bf
  x}-{\bf z}) J_0 ({\bf z}).
$
But the existence of the relation (\ref{eq:c_sym}), (\ref{eq:W_sym}),
(\ref{eq:new_sym}) saves this 
situation. Let us consider the transformation of the
condition (\ref{eq:c_sym}), (\ref{eq:W_sym}), (\ref{eq:new_sym}):
\begin{eqnarray*}
0=\int d^2 {\bf x}~ c_i ({\bf y}-{\bf x}) {\cal A}^i ({\bf x})
\rightarrow 
0&=& \int d^2 {\bf x}~ {c_i} ' ({\bf y}-{\bf x}) {{\cal A}^i} ' ({\bf x})
\nonumber \\  
&=&\int d^2 {\bf x}~ c_i({\bf y}-{\bf x}) {\cal A}^i ({\bf x})
-2 \int d^2 {\bf x}~ b ({\bf y}-{\bf x}) B({\bf z})
\nonumber \\
&\approx&  - \frac{2}{\kappa} \int d^2 {\bf x} ~b ({\bf
  y}-{\bf x}) J_0 ({\bf x})
\end{eqnarray*}
where we have neglected the boundary terms in the third line
for the rapidly decreasing functions $c_i ({\bf y}-{\bf
  x})$ and $b ({\bf y}-{\bf x})$. Then, one can find that 
  `$
  b ({\bf y}-{\bf x}) =0   
  $'
is the only solution for $b ({\bf y}-{\bf x})$ which retains the
condition (\ref{eq:identity}) when {\bf x} is located at the source
points where $J_0 ({\bf x})$ does not 
vanishes. It is interesting to note that the identity
(\ref{eq:identity}) 
corresponds to the divergence-free condition $\nabla \cdot {\bf A}^T =0$
for the gauge invariant transverse component ${\bf A}^T$ in
(\ref{eq:tl_decomp}): As the 
divergence-free condition defines ${\bf
  A}^T$, the condition (\ref{eq:identity}) can be also considered as a
defining 
equation for ${\cal A}_i$. On the other hand, it is easy to see that
only for the Coulomb gauge, the identity (\ref{eq:identity}) is
reduced to the 
divergence-free condition which is consistent to the fact of the
equivalence of ${\cal A}_i $ and $A_i$ in this gauge.

There is one more interesting effect of the condition
(\ref{eq:identity}). To see 
this, let us consider the condition (\ref{eq:identity}) with ${\cal
  A}_i $ expressed 
by the solution of (\ref{eq:AA_i_solution})
\begin{eqnarray}
\label{eq:identity_2}
0&=&\int d^2 {\bf x}~ c_i({\bf y}-{\bf x}) {\cal A}^i ({\bf x})
\nonumber \\
~&=& \int d^2 {\bf z}~ B ({\bf z}) \int d^2 {\bf x} ~\epsilon_{ki} c_i
({\bf y}-{\bf x}) c_k ({\bf z}-{\bf x}).
\end{eqnarray}
In general, the function $c_k ({\bf z}-{\bf x})$ is sum of the parity
even and odd parts. However, if one restricts only one part, i.e.,
$
c_i({\bf y}-{\bf x}) = \pm c_i({\bf x}-{\bf y})  
$
for the parity even or odd parts, respectively, (\ref{eq:identity_2}) becomes
\begin{eqnarray*}
  0&=& \mp \int d^2 {\bf z}~ B ({\bf z}) \Delta ({\bf z}-{\bf y} )
  \nonumber \\
  &\approx& \mp \frac{1}{\kappa} \int d^2 {\bf z}~ J_0 ({\bf z})
  \Delta ({\bf z}-{\bf y} ) 
\end{eqnarray*}
and one finds finally
\begin{eqnarray*}
  \Delta ({\bf z}-{\bf y} ) \approx 0
\end{eqnarray*}
when ${\bf z}$ or ${\bf y}$ is the position of the sources. [Here, the
former and latter coordinates ${\bf z}$ and ${\bf y}$, respectively have
no absolute meaning because of the symmetry
(\ref{eq:Delta_sym}). Moreover, in this 
case the coordinates in the function $ b ({\bf y}-{\bf x} )$ are also
equal footing because of the symmetry of (\ref{eq:c_sym}) which should be
preserved by $ b ({\bf y}-{\bf x} )$ (actually with opposite parity).]
This result provides an simple interpretation of the result
(\ref{eq:Delta=0}) for 
the solution (\ref{eq:c_solution_Coul}) which has odd parity. On the
other hand, the fact of 
$\Delta =0$ in (\ref{eq:Xi_axi}) even for the solution 
(\ref{eq:c_solution_axi}) which doesn't have the
definite parity is the result of the particular form of
(\ref{eq:c_solution_axi}): For more 
general solution (\ref{eq:c_axi_gen}), non-vanishing $\Delta$ is
expected.\footnote{$\Delta =0$ is retained under (\ref{eq:c_sym}) only
  for the case where $b ({\bf x}-{\bf y} )= b ({\bf y}-{\bf 
    x} ), ~ \int d^2 {\bf z} ~ \partial_j^z [ b ({\bf x}-{\bf
      z} ) c_j ({\bf y}-{\bf z} )- b ({\bf y}-{\bf z} ) c_j ({\bf
      x}-{\bf z} ) ]=0$ are satisfied; actually these are the condition
    for the invariance of $\Delta$ and $\xi$ in general. }

Now, finally we note that there is an identity for $A_k$-field as a
dual to (\ref{eq:identity}).  To this end, let us assume that
we can write $A_{\mu}$ as
\begin{eqnarray}
\label{eq:A}
  A_{\mu} ({\bf x}) = {\cal A}_{\mu} ({\bf x}) -\partial _{\mu}^x \int
  d^2 {\bf z}~ {\cal K}_k ({\bf x}-{\bf z} ) {\cal A}^k ({\bf x})
\end{eqnarray}
with
\begin{eqnarray}
\label{eq:K_dressing}
  \partial^{k}_{z} {\cal K}_{k}({\bf x}-{\bf z})=-\delta^2({\bf x}-{\bf z}).
\end{eqnarray}
Then, it is easy to show that $A_k$ in (\ref{eq:A})
satisfies\footnote{Note that $A_k$ does not gauge transform by the
  definition of (\ref{eq:A}): (\ref{eq:identity_A}) is just an
  identity for the completely 
  gauge fixed quantity $A_k$. In this sense (\ref{eq:identity_A}) is
  not the gauge 
  fixing condition which restricts the gauge variations: If we
  consider (\ref{eq:identity_A}) as a gauge fixing, the representation
of ${\cal K}_i$ as
  (\ref{eq:K_dressing}) is possible only for the Coulomb gauge as we
  have studied in 
  Sec. {\bf A}. (See \cite{Ban:93} for comparison) }
\begin{eqnarray}
\label{eq:identity_A}
\int d^2 {\bf z}~ {\cal K}_{k}({\bf x}-{\bf z}) A^{k} ({\bf z}) =0  
\end{eqnarray}
when ${\cal K}_{k}({\bf x}-{\bf z})$ is  rapidly
decreasing asymptotically such that the boundary term
$
  \int d^2 {\bf x} ~\partial_x^i ( {\cal K}_i ({\bf y}-{\bf x})  \int 
 d^2 {\bf z}~ {\cal K}_k ({\bf x}-{\bf z}) {\cal A}^k ({\bf z}) )
$
can be neglected. However, we note that there is no
direct connection between the gauge-fixing kernel ${\cal K}_k ({\bf x}
-{\bf z}) $ and the dressing function ${c}_k ({\bf x}-{\bf z})$ but
only the condition,
$
  \int d^2 {\bf x} ~\partial_x^i \left( {\cal K}_i ({\bf y}-{\bf x})
W({\bf x}) \right) =0
$
such that we obtain the gauge condition (\ref{eq:identity_A})
consistently. [This can be easily 
obtained by expressing ${\cal A}_{\mu}$ in the right-hand side of
(\ref{eq:A}) in terms of $A_{\mu}$ using (\ref{eq:phys_variable}) and
comparing the left and right- hand sides.] 

\begin{center}
{ \bf  V. Discussion and summary}
\end{center}

\begin{center}
{\bf  A. Gauge invariance in action}
\end{center}

In this paper, we have studied the manifestly gauge invariant
``Hamiltonian'' formulation where the energy momentum tensors 
are gauge invariant. But this does
not imply the gauge invariance of the action\footnote{Gauge invariance
of Lagrangian (density) is a requirement which is stronger than that
of action integral. Appreciation of this subtlety becomes necessary
recently, in discussions of the radiatively induced Lorentz and CPT
violating CS term in QED \cite{Jac:98}.}in general: The non-Abelian CS
gauge theory is an example \cite{Des:84,Des:82}. Hence, 
if we might find a representation corresponding to
(\ref{eq:phys_variable}) in the 
non-Abelian CS theory, it will be a problem to get the manifestly
gauge invariant action from the original action in terms of the
manifestly gauge invariant fields. However, in our case of Abelian
theory, there will be no general obstacle to do this. Actually it is found
that this is the case: When the gauge field $A_{\mu}$ in the CS term
of the original action (\ref{eq:CS_action}) is replaced by the gauge
invariant fields 
${\cal A}_{\mu}$, it reads
\begin{eqnarray*}
  \int d^3 x ~\frac{\kappa}{2} \epsilon^{\mu \nu \rho} {\cal A}_{\mu}
  \partial_{\nu} 
  {\cal A}_{\rho}&=&\int d^3 x ~\left[\frac{\kappa}{2}\epsilon^{\mu
  \nu \rho} {A}_{\mu}  
 \partial_{\nu}{ A}_{\rho}- \frac{\kappa}{2}\epsilon^{\mu \nu \rho} 
\partial_{\nu} {A}_{\rho} \partial_{\mu} W \right] \nonumber \\
&=& \int d^3 x ~\frac{\kappa}{2}\epsilon^{\mu \nu \rho} {A}_{\mu} 
 \partial_{\nu}{ A}_{\rho}-  \int d^3 x ~\frac{\kappa}{2}
  \partial_{\mu} \left[ 
\epsilon^{\mu \nu \rho} 
\partial_{\nu} {A}_{\rho} W \right] \nonumber \\
&\approx& \int d^3 x ~\frac{\kappa}{2}\epsilon^{\mu \nu \rho} {A}_{\mu} 
 \partial_{\nu}{ A}_{\rho}+ \int d^3 x ~\frac{1}{2} \partial_{\mu} \left[
 J^{\mu}  W \right] \nonumber \\
\end{eqnarray*}
; for the well-localized fields one finds the equivalence of the
original CS action of the base fields to the CS action of gauge invariant
field ${\cal A}_{\alpha}$. Furthermore, since the matter parts can be
made to be manifestly gauge 
invariant trivially, the total action integral is invariant
manifestly. Actually, in this derivation the coefficient $\kappa$ does
not have any role and actually this fact has been considered as the
signal of the no-quantization of $\kappa$ in a slightly different
context \cite{Des:84}. According to this interpretation, it is
expected that the
complete transformation of the original action into the action
which is expressed by the
manifestly gauge invariant fields in the non-Abelian CS theory, if it
does exist, at least up to the total
derivative terms, is not trivial
matter depending on the coefficient $\kappa$. If the interpretation is
a correct one, we suspect that 
\begin{eqnarray*}
 \int d^3 x ~ \kappa \epsilon^{\mu \nu \rho} \left< {\cal A}_{\mu}
  \partial_{\nu} 
  {\cal A}_{\rho} +\frac{2}{3} {\cal A}_{\mu} {\cal A}_{\nu} {\cal
  A}_{\rho} \right>  +8 \pi^2 \kappa \omega
=\int d^3 x ~\kappa \epsilon^{\mu \nu \rho} \left< { A}_{\mu} \partial_{\nu}
 {A}_{\rho} +\frac{2}{3} {A}_{\mu} {A}_{\nu} {A}_{\rho} \right>,
\end{eqnarray*}
[$\left< \cdots \right>$ denotes trace] up to the
total derivatives term,
where $\omega$ is an integer number which is involved to the winding
number for the homotopically non-trivial, large gauge transformations
of a non-Abelian gauge group whose $\pi_3 $ is {\bf Z}: Only in this
form, it is consistent with the well-known quantization of $\kappa$
\cite{Des:82}. 
But, it is unclear whether the manifestly gauge invariant variables
corresponding to (\ref{eq:phys_variable}) for all
gauge transformation, i.e., large as well as small gauge
transformations, exist or not.
\begin{center}
{\bf  B. Kinetic mass term and dual connection to Maxwell-CS theory}
\end{center}

In our original theory (\ref{eq:CS_action}) there is $U(1)$ gauge
symmetry and so the kinetic 
mass term $\frac{1}{2} \mu^2 A_{\mu}A^{\mu}$, which breaks the symmetry
manifestly, can not be introduced in this context. However, in
contrast, within the gauge invariant fields context, the mass like
term
\begin{eqnarray}
\label{eq:mass_1}
 \frac{1}{2} \mu^2 {\cal A}_{\mu}{\cal A}^{\mu} 
\end{eqnarray}
can not be discarded generally \cite{Park:97,Kas:94}. To see what this term
implies, let us re-express this term as follows, using the formulas in
the third steps in (\ref{eq:AA_i_solution}) and
(\ref{eq:AA_0_solution}) which are valid model independently 
\begin{eqnarray}
\label{eq:mass_2}
{\cal A}_{\mu}{\cal A}^{\mu} ({\bf x})
&=&\int d^2 {\bf z}\int d^2 {\bf y}~ c_k ({\bf x}-{\bf z})c_l ({\bf
  x}-{\bf y})
\left[ F_{k0}({\bf z}) F^{l0}({\bf y}) +F_{ki}({\bf z}) F^{li}({\bf
    y}) \right] \nonumber \\
&=&\int d^2 {\bf z}\int d^2 {\bf y}~ c_{\mu} ({\bf x}-{\bf z})c_{\nu} ({\bf
  x}-{\bf y}) F_{\mu \sigma}({\bf z}) F^{\nu \sigma}({\bf y})
\end{eqnarray}
with $c_0 ({\bf x}-{\bf z}) \equiv 0$, without using the equations of
motion. The resulted expression is similar to the Maxwell's kinetic
term,  but in 
a more generalized form, which is absent in the original theory
(\ref{eq:CS_action}). So, in order to introduce the mass term
(\ref{eq:mass_1}), we must also
consider an extension from the (pure) CS gauge theory (1). On the
other hand, since the result (\ref{eq:mass_2}) was obtained without using the
equation of motion, i.e., model independently, the extended theory
need not have exactly the same form as the final form of
(\ref{eq:mass_2}): These 
two theories may be related only after using the equations of motion
or the constraints. The most
natural candidate for the extended theory will be, of course, the
Maxwell-CS (MCS) theory 
which has the Maxwell term in addition to the CS theory (\ref{eq:CS_action})
\cite{Des:82,Des:84}:
\begin{eqnarray}
\label{eq:MCS}
{\cal L}_{MCS} [A]= -\frac{1}{4} F_{\mu \nu} F^{\mu \nu} + 
\frac{\kappa}{2} \epsilon^{\mu \nu \rho}A_{\mu}\partial_{\nu}A_{\rho}
+(D_{\mu}\phi)^{*}(D^{\mu}\phi)-m^{2} \phi^{*} \phi .
\end{eqnarray}
The theory has the equations of motion,
$
  \partial _{\mu}F^{\mu \nu} +\frac{\kappa}{2} \epsilon^{\mu \nu \rho}
  F_{\mu \rho} = J^{\nu}.
$
For our purpose, let us consider the special type of dressing
(\ref{eq:c_special}) and 
then the gauge invariant fields (\ref{eq:phys_variable}) become [here,
we start from the model independent formulas in
(\ref{eq:AA_i_solution}) and (\ref{eq:AA_0_solution})] 
\begin{eqnarray}
\label{eq:A_i_2}
  {\cal A}_i ({\bf x})&=&-\int d^2 {\bf z}~ c_k ({\bf x}-{\bf z}) F_{ki}
  ({\bf z}) \nonumber \\
&=&\int d^2 {\bf z} ~\chi ({\bf x}-{\bf z}) \partial_k^z F_{ki} ({\bf
  z}) \nonumber \\
&=&-\kappa \int d^2 {\bf z}~ \chi ({\bf x}-{\bf z}) \left( \epsilon^{ij}
  F_{j0}- \frac{1}{\kappa} \tilde{J}^i \right) ({\bf z}) ,  \\
\label{eq:A_0_2}
 {\cal A}_0 ({\bf x})&=&-\kappa \int d^2 {\bf z} c_k ({\bf x}-{\bf z}) F_{k0}
  ({\bf z}) \nonumber \\
&=&\int d^2 {\bf z} ~\chi ({\bf x}-{\bf z}) \partial_k^z F_{k0} ({\bf
  z}) \nonumber \\
&\approx&-\kappa \int d^2 {\bf z} ~\chi ({\bf x}-{\bf z}) \left( B +
  \frac{1}{\kappa} {J}^0 \right) ({\bf z}) ,   
\end{eqnarray}
where we have neglected the boundary term
$
 \int d^2 {\bf z}~\partial_k^z \left[ \chi ({\bf x}-{\bf z})  F^{k\mu} ({\bf
  z}) \right] . 
$
Here, we have introduced the ``convection current'',
$
  \tilde{J}^i \equiv J^i -\partial_0 F_{i0}
$
whose divergence is generated only from the CS part: 
$
  \nabla \cdot \tilde{\bf J} = \kappa \dot{B}
$
; the covariant form $\tilde{J}^{\mu} \equiv J^{\mu} -\partial_0
F_{\mu 0}$ is also available from $\tilde{J}^0=J^0$. Then, by using
(\ref{eq:A_i_2}) and (\ref{eq:A_0_2}), one
can express the mass term (\ref{eq:mass_2}) in terms of $F_{\mu \nu}$
as follows 
[we neglect the boundary term in the same ways as (\ref{eq:A_i_2}) and 
(\ref{eq:A_0_2})]
\begin{eqnarray*}
{\cal A}_{\mu}{\cal A}^{\mu}({\bf x}) &=&\kappa^2 \int d^2 {\bf z}
  \int d^2 {\bf 
  y}~ \chi({\bf x}-{\bf z}) \chi({\bf x}-{\bf y}) \left[ (B
  +\frac{1}{\kappa} J^0 )^2 -(F_{j0} -\frac{1}{\kappa} \epsilon^{lj}
  \tilde{J}^l )^2 \right] \nonumber \\
&=&\kappa^2 \int d^2 {\bf z}\int d^2 {\bf
  y} ~\chi({\bf x}-{\bf z}) \chi({\bf x}-{\bf y}) \left[ (B^2
  -F_{j0}F_{j0} )-\frac{4}{\kappa} \epsilon^{\mu \nu \rho} F_{\nu \rho}
  \tilde{J}_{\mu} +\tilde{J}^{\mu} \tilde{J}_{\mu} \right] \nonumber\\
&=&\frac{\kappa^2}{2} \int d^2 {\bf z} \int d^2 {\bf
  y} ~\chi({\bf x}-{\bf z}) \chi({\bf x}-{\bf y}) F_{\mu \nu}({\bf z})
F^{\mu \nu}({\bf y}) \nonumber \\
&& +  (\tilde{J}_{\mu}-\mbox{dependent~ terms}).
\end{eqnarray*}
The final form of the right-hand side looks
like the Maxwell term of (\ref{eq:MCS}), but it is still different by
the non-local 
expression through $\chi$-functions. Moreover, because of the explicit
appearance of the function $\chi$ which is absent in the Lagrangian
(\ref{eq:MCS}), this term has no counter parts in (\ref{eq:MCS}). In
order to resolve 
these problems, let us consider $\Box {\cal A}_{\mu}$ instead of
${\cal A}_{\mu}$: 
\begin{eqnarray}
\label{eq:box_A}
\Box {\cal A}_0 ({\bf x}) &=&-\kappa \int d^2 {\bf z}~
\Box^x \chi ({\bf x}-{\bf z})\left(B +\frac{1}{\kappa} J^0 \right)
({\bf z})
-\kappa \int d^2 {\bf z}~
\partial^z_k \left(\Box^x \chi ({\bf x}-{\bf z}) F_{k0} ({\bf z})
\right) \nonumber \\
&=&\kappa \left( B +\frac{1}{\kappa} J^0 \right) ({\bf x}) + \int d^2
{\bf z} ~\partial^z_k \left( \delta^2 ({\bf x}-{\bf y}) F_{k0} ({\bf
    z}) \right), \nonumber \\
\Box {\cal A}_i ({\bf x}) &=&-\kappa \int d^2 {\bf z}~
\Box^x \chi ({\bf x}-{\bf z})\left( \epsilon^{ij} F_{j0}
  +\frac{1}{\kappa} \partial_0 F_{i0} -\frac{1}{\kappa} J^i \right)
-\kappa \int d^2 {\bf z}~
\partial^z_k \left(\Box^x \chi ({\bf x}-{\bf z}) F_{ki} ({\bf z})
\right) \nonumber \\
&=&\kappa \left( \epsilon^{ij} F_{j0}-\frac{1}{\kappa} \tilde{J}^i
\right) ({\bf x}) + \int d^2 
{\bf z} ~ \partial^z_k \left( \delta^2 ({\bf x}-{\bf y}) F_{ki} ({\bf
    z}) \right).
\end{eqnarray}
Then one obtains
\begin{eqnarray}
\label{eq:box_mass_1}
  -\frac{1}{2 \kappa^2} \Box {\cal A}_{\mu} \Box {\cal
   A}^{\mu}&=&-\frac{1}{2} (B +\frac{1}{\kappa} J^0 )^2 +\frac{1}{2}
   (F_{j0} -\frac{1}{\kappa} \epsilon^{lj} \tilde{J}^l )^2 \nonumber
   \\
&=&-\frac{1}{4} F_{\mu \nu} F^{\mu \nu} + (\tilde{J}^{\mu}-\mbox{dependent
  ~terms} ), 
\end{eqnarray}
where we have neglected the singular boundary terms in (\ref{eq:box_A}) and
`$\tilde{J}^{\mu}$-dependent terms' is
$
  \frac{1}{2 \kappa} \epsilon^{\mu \nu \rho} \tilde{J}^{\mu} F_{\nu
  \rho} -\frac{1}{2 \kappa ^2} \tilde{J}^{\mu}\tilde{J}_{\mu} .
$
On the other hand, by noting the wave equation for $F^{\mu \nu}$ [31],
$
  (\Box + \kappa^2 ) F^{\mu \nu} = \kappa \epsilon^{\mu \nu \rho}
  J_{\rho} + \partial^{\mu} J^{\nu} -\partial^{\nu} J^{\mu}
$
and so the wave equation for the gauge invariant fields, before using
(\ref{eq:c_special}), 
\begin{eqnarray*}
\Box {\cal A}_{\mu} ({\bf x}) &=&\int d^2 {\bf z}~
\Box^x c_k ({\bf x}-{\bf z}) F_{k \mu} ({\bf z}) \nonumber \\
 &=&-\int d^2 {\bf z}~
c_k ({\bf x}-{\bf z}) \Box^{z}  F_{k \mu} ({\bf z}) \nonumber \\  
&=&-\kappa^2 {\cal A}_{\mu} ({\bf x})-\int d^2 {\bf z}~
c_k ({\bf x}-{\bf z})\left( \kappa \epsilon^{k \mu \rho} J_{\rho} +
  \partial^k J^{\mu} -\partial^{\mu} J^k \right) ({\bf z})
\end{eqnarray*}
[we have neglected the boundary terms in the second line] one finds
that the left-hand side of (\ref{eq:box_mass_1})
becomes the usual mass term beside the $J^{\mu}$-dependent terms
\begin{eqnarray}
\label{eq:box_mass_2}
  -\frac{1}{2 \kappa^2} \Box {\cal A}_{\mu} \Box {\cal A}^{\mu}=
-\frac{1}{2 \kappa^2} {\cal A}_{\mu} {\cal A}^{\mu}
+(J^{\mu}-\mbox{dependent~term}).
\end{eqnarray}
Hence, by combining (\ref{eq:box_mass_1}) and (\ref{eq:box_mass_2})
together, the mass 
term (\ref{eq:mass_1}) corresponds to
the Maxwell term in the MCS when we neglect both $J^{\mu}$ and
$\tilde{J}^{\mu}$-dependent terms and the singular boundary terms. On
the other hand, we note that the 
neglected $J^{\mu}$ and $\tilde{J}^{\mu}$-dependent terms and the
singular boundary terms are order of
$\frac{1}{\kappa}$ with respect to the first terms in
(\ref{eq:box_mass_1}) and 
(\ref{eq:box_mass_2}). Hence, it is found that the
so-called ``self-dual'' action \cite{Tow:84,Des:84} with respect to the gauge
invariant fields ${\cal A}_{\mu}$
\begin{eqnarray*}
  I_{SD}[{\cal A}] =\int d^3 x ~\left[\frac{\kappa}{2} \epsilon^{\mu
  \nu \rho} {\cal 
  A}_{\mu} \partial_{\nu} {\cal A}_{\rho}-\frac{\kappa^2}{2} 
{\cal A}_{\mu}{\cal A}^{\mu} \right]
\end{eqnarray*}
corresponds to the MCS theory (or topologically massive gauge
theory \cite{Des:82} ) without ``dynamical matter'' parts when we neglect the
${\cal O} (\frac{1}{\kappa})$ terms:
\begin{eqnarray*}
  I_{SD}[{\cal A}] =I_{MCS} [A] +{\cal O} (\frac{1}{\kappa})
\end{eqnarray*}
; the result is similar to the previous work
\cite{Des:84,Ban:95} in that the SD and MCS theory
are equivalent only when we neglect the dynamical matters\footnote{It
  is not understood why the two different types of currents are
  involved in SD and MCS sides each other.}, though
equivalent even with the {\it external currents}; but the discrepancy of the
order of ${\cal O} (\frac{1}{\kappa})$, which looks like a pertubative
correction in the path-integral approach of a model \cite{Frd:94}, is not
understood. Moreover, we have considered the special case of 
(\ref{eq:c_special}) in
the proof and it is unclear whether the similar equivalence can be
proved in more general cases or not; this special case of
(\ref{eq:c_special}) which is involved 
with the Coulomb gauge in GFF, might be connected to the results of
the {\it phase space} path-integral approach of Ref. \cite{Ban:95},
where the Coulomb gauge was crucial for the equivalence of Lagrangians,
but complete connection is not known\footnote{In the configuration
  space analysis, in contrast, the Lorentz gauge is crucial for the
  equivalence \cite{Des:84,Des:82}; in this case, the Lorentz gauge in the
  self-dual frame is nothing but the Bianchi identity in the MCS
  frame.}. Finally, we note that our method provides a new framework
for finding 
the corresponding dual theory compared to the previously known methods
\cite{Des:84,Ban:95,Ban:96,Wit:95}.

In summary of this paper, we have considered a new GIF consistent with
GFF.  Our 
formalism is new in the following three points. (A) We introduced the
assumption that there be no translation transformation anomaly for
gauge invariant variables ${\cal F}_{\alpha}$. From this assumption,
we obtained several new conditions for the dressing function
$c_{k}({\bf x}, {\bf z})$, which are crucial in our development. (B)
We introduced the master formula (\ref{eq:master_eq}), which allowed
matching to the 
gauge fixed system. (C) We found the manner how the equation
of the dressing function $c_{k}({\bf x}, {\bf z})$ are modified after
gauge fixing. Using this formulation, we have obtained a novel GIF,
which is consistent with the conventional GFF: The former formulation
provides exactly the rotational anomaly of the latter.  Hence, in our
formulation there is no inconsistency, as in the previous gauge
independent formulation of Ref. \cite{Ban:92}.  As a byproduct, we explicitly
found that the anomalous spin of the charged matter has a unique
meaning. This is due to the uniqueness of the Poincar\'e
generators when constructed from the symmetric energy-momentum
tensor because it is the improved generator which are
(manifestly) gauge invariant and obey the quantum as well as classical
Poincar\'e algebra, but this is not the case for the canonical
Poincar\'e generators. Moreover, we have constructed the physical
states in the algebraic construction and also in the Schr\"odinger
picture. In 
the latter method, we found that the ``gauge invariant'' scalar field
$\eta$ of the 
longitudinal mode of $A_i$ is crucial for constructing the physical
wavefunctional which is a genuine effect of (pure) CS theory. The
existing confusion about the gauge condition and dressing function
have been clarified.

We would like to conclude with two additional comments. First, in
our formulation, there is no gauge non-invariance problem of
Poincar\'e generators on the physical states.  This is essentially due
to absence of additional terms proportional to constraints in the
generators of (\ref{eq:Poincare_gen}), in contrast to the old
formulation of Dirac 
\cite{Dir:55}.
Second, the master formula (\ref{eq:master_eq}), which guarantees the
classical 
Poincar\'e covariance of our CS gauge theory in all gauges, also works
in all other gauge theories.  Hence as far as the $gauge~ dependent$
operator ordering problem does not occur, the $quantum$ Poincar\'e
covariance for one gauge guarantees also the covariance for all other
gauges. The gauge independent proof of quantum covariance has been an
old issue in quantum field theory, and now it is reduced to the
solvability of the problem of the gauge dependent operator ordering.

\begin{center}
  { \bf Acknowledge}
\end{center} 

One of us (M.-I. Park) would like to thank Prof. Roman Jackiw for
reading a draft paper of this work and giving several 
valuable comments and suggestions, and
Profs. Doochul Kim, Choonkyu Lee, Jae Hyung Lee, and Hee Sung Song for warm
hospitality of providing a financial support and researching
facilities. He also thank Prof. Stanley Deser, Drs. Chanju
Kim, Yong-Wan Kim, and Hyun Seok Yang for discussions. This work was 
supported in part by the Korea Science and Engineering Foundation
(KOSEF), Project No. 97-07-02-02-01-3 and the Korea Research
Foundation (KRF), Project No. 1998-015-D00074.

\begin{center}
  { \bf Appendix A. Dirac's extended Poincar\'e generators}
\end{center}

In this Appendix, we will consider the extension of the generators
(\ref{eq:Poincare_gen}) by including the constraints terms and we will
show that the 
correct transformation law can be obtained for the undressed base
fields $F_{\alpha}=(A_{\mu}, \phi, \phi^* )$ 
as well as the gauge invariant fields ${\cal F}_{\alpha}=({\cal
  A}_{\mu}, \hat{\phi}, \hat{\phi}^* )$. This method 
has been widely used after the formulation by Dirac
\cite{Ban:92,Dir:64,Hen:92}
and has been
considered as what having a general validity. But in this appendix we will
show that this is valid only when we neglect the singular boundary
terms \cite{Park:97}.

To this end, let us consider the extended generators by the
first-class constraints $T \approx 0$ and $T_0 \approx 0$ [here, we
included the constraint $T_0 \approx 0$ in order to give the correct
transformation to $A_0$ also] \cite{Dir:55,Dir:64}
\begin{eqnarray} 
\label{eq:Poincare_gen_ex}
&&{P}^{0}_{s(E)} =\int d^2 {\bf x} ~ \left[ |\pi_{\phi}|^{2} +|{D}^{i}
  \phi|^{2}  
+m^{2} |\phi|^{2}+{v}^{0}_{0} T_0+ v^0 T \right], \nonumber \\
&&{P}^{i}_{s(E)} =\int d^{2} {\bf x }~\left[\pi_{\phi} D^{i}\phi +({D}^{i}
\phi)^{*} \pi_{\phi}^{*}
+{v}^{i}_{0} T_0+ v^i T \right], \nonumber \\
&&M^{12}_{s(E)} = \int d^2 {\bf x} ~ \left[ \epsilon_{ij}x^{i}
\left(\pi_\phi D^j \phi +(D^j \phi)^* \pi_\phi^* \right)
+u_0 T_0+ u T \right], \nonumber \\
&&{M}^{0i}_{s(E)} = x^{0} P^{i}_{s}-
\int d^{2} {\bf x} ~ \left[x^{i}\left(|\pi_{\phi}|^{2} +|D^{j}
  \phi|^{2} +m^{2} |\phi|^{2} \right) 
+{u}^{i}_{0} T_0+ u^i T \right],
\end{eqnarray}
and consider the transformation property of the undressed base
fields ${A}_{\mu},~\phi,~\phi^{*}$; the transformation for the gauge
invariant fields are the same as in Sec. II.

Firstly, for the time translational generator ${P}^{0}_{s(E)}$, this
produces the following transformation
\begin{eqnarray*}
&&\{{A}_{0}({\bf x}), {P}^{0}_{s(E)} \} \approx {v}^{0}_{0}
({\bf x}), \nonumber \\ 
&&\{{A}_{i}({\bf x}), {P}^{0}_{s(E)} \} \approx  \epsilon_{ij}
\frac{1}{\kappa}{J}_{j}({\bf x})+\partial^{i} v^{0}({\bf x}),
\nonumber \\ 
&&\{{\phi}({\bf x}), {P}^{0}_{s(E)} \} \approx
\pi_{\phi}^{*}({\bf x})+i v^0 \phi ({\bf x}), \nonumber \\
&&\{{\phi}^{*}({\bf x}), {P}^{0}_{s(E)} \} \approx \pi_{\phi}({\bf
  x})-i v^0 \phi^{*}({\bf x}) 
\end{eqnarray*}
such that ${P}^{0}_{s(E)}$ can be made to produce the correct
transformation
$
\{{F}_{\alpha}({\bf x}), {P}^{0}_{s(E)} \} \approx \partial^{0}
{F}_{\alpha}({\bf x}),~~~~~~~~~~~~
$
if the coefficients are
\begin{eqnarray}
\label{eq:v^0}
&&{v}^{0}_{0} \approx \partial^{0}A_{0}, \nonumber \\
&&\partial_{i} {{v}}^{0} \approx \epsilon_{ij}\frac{1}{\kappa}
{J}_{j}-\partial_{0} {A}_{i}, \nonumber \\ 
&&v^0 \approx - A^{0}.
\end{eqnarray}

Note that these solutions are consistent with the classical equation
of motion ${F}^{0k}=\frac{1}{\kappa} \epsilon_{kj} {J}^j$ which
can be reproduced by combining the second and the third equations in 
(\ref{eq:v^0}) and the fact of $ \pi_{\phi}=(D_{0} \phi)^{*}$.

Secondly, for the space translation generators ${P}^{i}_{s(E)}$, this
produces the transformations
\begin{eqnarray*}
&&\{{A}_{0}({\bf x}), {P}^{j}_{s(E)} \} \approx {v}^{j}_{0}
({\bf x}), \nonumber \\
&&\{{A}_{i}({\bf x}), {P}^{j}_{s(E)} \} \approx  -\epsilon_{ij} 
\frac{1}{\kappa}{J}_{0}({\bf x})+\partial_{i} v^{j}({\bf x}), \nonumber \\
&&\{{\phi}({\bf x}), {P}^{j}_{s(E)} \} \approx
\left[(\partial^{j}+i{A}^{j}) +i {v}^{j} \right] \phi({\bf x}), \nonumber \\ 
&&\{{\phi}^{*}({\bf x}), {P}^{j}_{s(E)} \} \approx
\left[(\partial^{j}-i{A}^{j}) -i {v}^{j} \right] \phi^{*}({\bf x}) 
\end{eqnarray*}
such that ${P}^{j}_{s(E)}$ can be made to produce the desired
transformation
$
\{{F}_{\alpha}({\bf x}), {P}^{j}_{s(E)} \} \approx \partial^{j}
{F}_{\alpha}({\bf x}) 
$
if the coefficients are
\begin{eqnarray*}
&&{v}^{i}_{0} \approx \partial^{i}A_{0}, \nonumber \\
&&\partial_{i} {{v}}^{j} \approx -\epsilon_{ij}\frac{1}{\kappa}
{J}_{0}+\partial_{j} {A}_{i}, \nonumber \\ 
&&{v}^{j} \approx - {A}^{j}.
\end{eqnarray*}
Note that these solutions are also consistent with the constraint
$B=\frac{1}{\kappa} J_0$.

Thirdly, for the space-rotation generator ${M}^{12}_{s(E)}$, this
produces the transformations
\begin{eqnarray*}
&&\{{A}_{0}({\bf x}), {M}^{12}_{s(E)} \} \approx {u}_{0} ({\bf
  x}), \nonumber \\
&&\{{A}_{i}({\bf x}), {M}^{12}_{s(E)} \} \approx  x_{i} 
\frac{1}{\kappa}J_{0}({\bf x})-\partial_{i} u({\bf x}) + \int d^2 {\bf
  z} ~ \partial^z_i \left( u ({\bf z}) \delta^2 {\bf x} -{\bf z})
  \right) , \nonumber \\ 
&&\{{\phi}({\bf x}), {M}^{12}_{s(E)} \} \approx
  \left[\epsilon_{ij}x^{i}(\partial^{j}+i{A}^{j}) +i u \right] \phi({\bf x}),
  \nonumber \\ 
&&\{ \phi^*({\bf x}), M^{12}_{s(E)} \} \approx
  \left[\epsilon_{ij}x^{i}(\partial^{j}-i{A}^{j}) -i u \right] \phi^{*}({\bf
  x}) 
\end{eqnarray*}
and because of the {\it singular} boundary term for $\{ A_i ({\bf
  x}), M^{12}_{s(E)} \}$, it is impossible to get the 
desired transformation
\begin{eqnarray*}
\{ F_{\alpha}({\bf x}), M^{12}_{s(E)} \}= (x^1 \partial^2
 -x^{2} \partial^1)F_{\alpha}({\bf x}) +\Sigma^{12}_{\alpha
   \beta} F_{\beta}({\bf x}), 
\end{eqnarray*}
for all base fields $F_{\alpha}$ unless we exclude the boundary positions in
the field point ${\bf x}$. In the case when $\phi$ (or $\phi^*$) has
no anomalous transformation, the coefficients are found to be
\begin{eqnarray}
\label{eq:u}
u_0 \approx \epsilon_{ij}x^i \partial^j A^0, ~~u \approx
-\epsilon_{ij}x^i A^j 
\end{eqnarray}
with the transformations
\begin{eqnarray*}
&&\{{A}_{0}({\bf x}), {M}^{12}_{s(E)} \} \approx
\epsilon_{jk}x^{j} \partial^{k} {A}_{0} ({\bf x}), \nonumber \\ 
&&\{{A}_{i}({\bf x}), {M}^{12}_{s(E)} \} \approx
\epsilon_{jk}x^{j} \partial^{k}{A}_{i}({\bf x})-\epsilon_{ik}
{A}_{k}({\bf x}) + \int d^2 {\bf z} ~ \partial^z_i \left
  ( \epsilon_{jk} z_j A^k ({\bf z}) \delta^2 ({\bf x}-{\bf z})
\right), \nonumber \\  
&&\{{\phi}({\bf x}), {M}^{12}_{s(E)} \} \approx
\epsilon_{jk}x^{j}\partial^{k} \phi({\bf x}), \nonumber \\ 
&&\{{\phi}^{*}({\bf x}), {M}^{12}_{s(E)} \} \approx
\epsilon_{jk}x^{j}\partial^{k} \phi^{*}({\bf x}).  
\end{eqnarray*}
It seems that there is no other solution which is better than
(\ref{eq:u}). It is not clear how this additional boundary term is related to
the anomaly for the gauge invariant fields ${\cal F}_{\alpha}$.

Finally, to consider the transformations generated by the
Lorentz-boost generators 
${M}^{0i}_{s(E)}$, it is more convenient to re-express
${M}^{0i}_{s(E)}$ in (\ref{eq:Poincare_gen_ex}) as
\begin{eqnarray*}
M^{0i}_{s(E)} = x^0 P^i_{s(E)}-
 \int d^{2}{\bf x}~ \left[ x^{i} \left(|\pi_{\phi}|^{2} +|D^{j}
 \phi|^{2} +m^{2} |\phi|^2 \right) 
+u^{'i}_0 T_0+ u^{'i}_3 T \right], 
\end{eqnarray*}
and this produces the transformations
\begin{eqnarray*}
&&\{{A}_{0}({\bf x}), {M}^{0j}_{s(E)} \} \approx x^{0}
\partial^{j}A_{0}({\bf x})-{u}^{'j}_{0}({\bf x}), \nonumber \\ 
&&\{{A}_{i}({\bf x}), {M}^{0j}_{s(E)} \} \approx
(x^{0}\partial^{j}-x^{j} \partial^{0}){A}_{i}({\bf
  x})-\delta_{ij}A_{0}({\bf x})+\partial_{i}(x^{j}A_{0}+
{u}^{'j} )({\bf x}) \nonumber \\
&&~~~~~~~~~~~~~~~~~~~~  -\int d^2 {\bf z} ~ \partial^z_i \left
  ( u^{'j}({\bf z}) \delta^2 ({\bf x}-{\bf y}) \right), \nonumber \\ 
&&\{{\phi}({\bf x}), {M}^{0j}_{s(E)} \} \approx (x^{0}
\partial^{j}-x^{j} \partial^{0}) \phi({\bf x}), \nonumber \\ 
&&\{{\phi}^{*}({\bf x}), {M}^{0j}_{s(E)} \} \approx (x^{0}
\partial^{j}-x^{j} \partial^{0}) \phi^{*}({\bf x}). 
\end{eqnarray*}
Similar to the rotation transformation, there is a solution for the
coefficients 
\begin{eqnarray*}
{u}^{'i}_{0}\approx x^{i}\partial^{0}A_{0}+{A}_{i},
~~{u}^{'i}\approx -x^{i}A_{0}.
\end{eqnarray*}
which produce the desired transformations
\begin{eqnarray*}
\{ {F}_{\alpha}({\bf x}), {M}^{0j}_{s(E)} \}= (x^{0} \partial^{j}
 -x^{j} \partial^{0}){ F}_{\alpha}({\bf x}) +\Sigma^{0j}_{\alpha
   \beta} {F}_{\beta}({\bf x}), 
\end{eqnarray*}
for the matter field $F_{\alpha}$ as follows
\begin{eqnarray*}
&&\{{A}_{0}({\bf x}), {M}^{0j}_{s(E)} \} \approx
(x^{0}\partial^{j}-x^{j}\partial^{0})A_{0}({\bf x})-{A}_{j}({\bf
  x}), \nonumber \\ 
&&\{{A}_{i}({\bf x}), {M}^{0j}_{s(E)} \} \approx
(x^{0}\partial^{j}-x^{j}\partial^{0}){A}_{i}({\bf
  x})-\delta_{ij}{A}_{0}({\bf x}) +\int d^2 {\bf z} ~ \partial^z_i \left
  ( z^j A_0 ({\bf z}) \delta^2 ({\bf x}-{\bf z}) \right), \nonumber \\ 
&&\{{\phi}({\bf x}), {M}^{0j}_{s(E)} \} \approx
(x^{0}\partial^{j}-x^{j}\partial^{0})\phi({\bf x}), \nonumber \\ 
&&\{{\phi}^{*}({\bf x}), {M}^{0j}_{s(E)} \} \approx
(x^{0}\partial^{j}-x^{j}\partial^{0})\phi^{*}({\bf x}).
\end{eqnarray*}

In conclusion, the Dirac's idea, which introduces the extended
Poincar\'e generator as a correct generator without choosing the gauge
condition, can be applied in the CS theory only when we neglect the
singular boundary terms. This will be the first example of this
phenomena as far as we know.
\begin{center}
  { \bf Appendix B}
\end{center}

In this Appendix, we explain the usual case where $W({\bf x})$ itself can be
diagonalized and the physical states can be constructed without
recourse to $\eta$ field.

Our considering model is the Maxwell-CS theory [the action will be
described in (\ref{eq:MCS})] and they produce the  Gauss' law constraint
\begin{eqnarray}
\label{eq:MCS_Gauss}
  \label{eq:B1}
  \left( \partial_i E^i + \kappa B - :J^0: \right) \left| \Psi_{\mbox{phys}}
  \right> =0
\end{eqnarray}
where $E^i =F^{i0}$. The non-vanishing commutation relations are
$
  [ A_i ({\bf x}), E^j ({\bf y}) ] = i \hbar \delta_{ij} \delta^2
  ({\bf x}-{\bf y})
$
and so, one can consider the representation where $A_i$ is
diagonalized:
\begin{eqnarray}
\label{eq:MCS_rep}
&&E^i ({\bf x}) \left| \Psi \right> \rightarrow \frac{\hbar}{i}
  \frac{\delta}{\delta A_i ({\bf x})} \Psi (A), \nonumber \\
&&A_i ({\bf x}) \left| \Psi \right> \rightarrow A_i ({\bf x}) \Psi (A).
\end{eqnarray}
Then, together with the representation for the matter parts
(\ref{eq:matter_rep}), the 
representation (\ref{eq:MCS_rep}) make the
Gauss' law (\ref{eq:MCS_Gauss}) become a differential equation
\begin{eqnarray}
  \label{eq:MCS_Gauss_1}
  \left[ \frac{\hbar}{i} \partial_i \frac{\delta}{\delta A_i ({\bf x})}
  + \kappa B ({\bf x}) - \hbar \phi ({\bf x}) \frac{\delta}{\delta
  \phi ({\bf x})} 
+\hbar \phi^{\dagger} ({\bf x}) \frac{\delta}{\delta \phi^{\dagger} ({\bf
  x})} \right] \Psi_{\mbox{phys}} =0.
\end{eqnarray}
On the other hand, since the first part becomes
\begin{eqnarray*}
  \label{eq:B5}
  \partial_i  \frac{\delta}{\delta A_i ({\bf x})} &=&\int d^2 {\bf y}~
  \left[ \partial^x_i \left( \frac{\delta {\cal A}_j ({\bf y})}{\delta
  A_i({\bf x})}\right) \frac{\delta}{\delta {\cal A}_j ({\bf y})} 
+\partial^x_i \left( \frac{\delta {W} ({\bf y})}{\delta
  A_i({\bf x})}\right) \frac{\delta}{\delta W({\bf y})} \right] 
  \nonumber \\
  &=&-\frac{1}{i} \frac{\delta}{\delta W ({\bf x})} 
\end{eqnarray*}
from the relation
\begin{eqnarray*}
 &&\partial^x_i \left( \frac{\delta {\cal A}_j ({\bf y})}{\delta
  A_i({\bf x})}\right)=\partial^x_j \delta^2 ({\bf x}-{\bf
  y})+\partial^y_j \delta^2 ({\bf x}-{\bf y})=0, \nonumber \\ 
 &&\partial^x_i \left( \frac{\delta {W} ({\bf y})}{\delta
  A_i({\bf x})}\right) =-\partial^x_i c_i({\bf y}-{\bf x})= -\delta^2
  ({\bf y}-{\bf x}) 
\end{eqnarray*}
(\ref{eq:MCS_Gauss_1}) becomes, finally
\begin{eqnarray}
  \label{eq:MCS_Gauss_2}
  \left[ -\frac{\hbar}{i} \frac{\delta}{\delta W ({\bf x})} + \kappa
  B({\bf x}) -\hbar \phi ({\bf x}) \frac{\delta}{\delta \phi ({\bf x})} +
\hbar \phi^{\dagger} ({\bf x}) \frac{\delta}{\delta \phi^{\dagger} ({\bf
  x})} \right] \Psi_{\mbox{phys}} =0.
\end{eqnarray}
Then, it is easy to see that the solution of (\ref{eq:MCS_Gauss_2}) is
the form of \cite{Des:82} 
\begin{eqnarray}
  \label{eq:MCS_phys_wave}
  \Psi_{\mbox{phys}} (B, W, \phi, \phi^{\dagger} ) =e^{- i \kappa \int
  d^2 {\bf x}~
  B({\bf x}) W ({\bf x})} \Phi (B) \varphi(\hat{\phi}, \hat{\phi}^{\dagger}),
\end{eqnarray}
when $\Phi (B)$ and $\varphi(\hat{\phi}, \hat{\phi}^{\dagger})$ are
any functionals of $B$ and $\hat{\phi},~ \hat{\phi}^{\dagger}$
respectively. Because of the first exponential factor, the physical
wavefunctional are not gauge invariant and actually the factor
produces the $1-cocyle$ for the MCS theory. Furthermore, we note that
$\eta$ field commutes with $W$ and 
$\partial_i E^i$, the $\eta$ part of $\Psi_{\mbox{phys}}$ is not
determined in any way and only has a 
redundant role. Finally, we note that this model has a well defined
weak coupling limit $\kappa \rightarrow 0$, i.e, the
three-dimensional QED limit and the wavefunctional is found to be
\begin{eqnarray*}
  \Psi_{\mbox{phys}} (B, W, \phi, \phi^{\dagger} ) = \Phi (B)
  \varphi(\hat{\phi}, \hat{\phi}^{\dagger}) 
\end{eqnarray*}
from (\ref{eq:MCS_phys_wave}); however, the strong coupling limit
$\kappa \rightarrow \infty$, i.e., pure CS limit is not well-defined.
\begin{center}
  {\bf Appendix C. Proof of master formula }
\end{center}

In this Appendix, we present the proof of the master formula
(\ref{eq:master_eq}). There are two class of the constraints system
largely: The 
first-class constraint system and the second-class constraint
one. In the former, one choose the gauge conditions to remove the
redundancies which makes the quantum theory is well-defined; in the
latter one does not introduce additional (gauge) conditions since
the quantum theory is well-defined (at least formally) without
it. These two systems are closely related but the exact equivalence is
generally unclear. So, for the definiteness, we will consider these
two cases separately.

First, let us consider the former case, i.e., the case when there is a
first-class constraints systems $T\approx 0$ and corresponding gauge
condition $\Gamma \approx 0$:
$
 \{ T, T \} \sim T \approx 0, 
  \{ T, \Gamma \} \not\approx 0.
$
[Here, it does
not matter what the bracket algebra for $\Gamma$ itself $\{\Gamma,
\Gamma \}$ is.] Then, the Poisson bracket matrices $\Delta_{\alpha \beta}
\equiv \{ \Theta_{\alpha}, \Theta_{\beta} \}~( \Theta_1 \equiv T
\approx 0, \Theta_2 \equiv \Gamma
\approx 0)$ becomes
\begin{eqnarray*}
  \Delta_{\alpha \beta} =\left( \begin{array}{cc}
                                             a &    b    \\
                                             c &    d
                                 \end{array}
                          \right), 
\end{eqnarray*}
where $a=\{ T, T \} \approx 0, 
                     b=\{ T, \Gamma \} \not\approx 0, c= \{ \Gamma , T \}
                     \not\approx 0$, and $ 
                     d= \{ \Gamma, \Gamma \}$ 

with non-vanishing determinant:
$
  det \Delta _{\alpha \beta} = a d -b c \approx - b c.
$
[Here, we are considering only the discrete indices for
convenience. Generalization to the continuous indices will be
straightforward.] 
Then, the inverse of $\Delta_{\alpha \beta}$ becomes
\begin{eqnarray*}
  \Delta^{-1} &=&\frac{1}{det \Delta } \left( \begin{array}{cc}
                   d & -b \\
                   -c & a  \end{array} \right) \nonumber \\
        &\approx& -\frac{1}{b c} \left( \begin{array}{cc}
                       d & -b \\
                       -c & 0 \end{array} \right) .
\end{eqnarray*}
Here, the fact of $\Delta^{-1}_{22} \approx 0$ is crucial in the
proof. Then, the Dirac bracket \cite{Dir:64}
for the gauge invariant $L_a$,
which has a vanishing Poisson bracket with respect to the first-class
constraints, i.e., $\{L_a, T \} \approx 0$ is found to be
\begin{eqnarray*}
  \{ L_a, L_b \}_{D_{\Gamma}} &\equiv& \{ L_a, L_b \} -\{ L_a,
  \Theta_{\alpha} \} 
  \Delta^{-1}_{\alpha \beta} \{ \Theta_{\beta}, L_b \} \nonumber \\
  &\approx& \{L_a, L_b \} -\{ L_a, \Theta_{2} \}
  \Delta^{-1}_{22} \{ \Theta_{2}, L_b \} \nonumber \\
  &\approx& \{L_a, L_b \}, 
\end{eqnarray*}
where we have used the conditions $\{ L_a, T \} \approx 0$ and
$\Delta^{-1}_{22} \approx 0$ in the second and third lines,
respectively. This proof can be generalized to the case of several
first-class constraints and corresponding gauge conditions; even with
the (partial) gauge conditions which are involved only for a part
of the first-class constraints, this proof is applied for the
partially gauge invariant variables which commute only with those
parts of the first-class constraints. This proves the master formula
(\ref{eq:master_eq}) for the former case.

Secondly, let us consider the later case, i.e., the case when there is
only one second-class constraint $\chi \approx 0$:
$
  \Delta \equiv \{ \chi, \chi \} \neq 0.
$
Then, the Dirac bracket for the quantity $L_a$, which commutes with
the second-class constraint $\chi$, i.e., $\{L_a, \chi \} \approx 0$ is
found to be 
\begin{eqnarray*}
  \{ L_a, L_b \}_D &=& \{L_a, L_b \} -\{ L_a, \chi \}
  \Delta^{-1}_{22} \{ \chi, L_b \} \nonumber \\
  &\approx& \{L_a, L_b \}, 
\end{eqnarray*}
where we have used the fact of $\{L_a, \chi \} \approx 0$. This proof
can be generalized to the case of 
several second-class constraints together with the first-class
constraints which do not have the involved gauge conditions when $L_a$
commutes with the second-class constraints. This latter case is rather
unusual case compared to the former case \cite{Hen:92}
and has been studied only recently in the CS theories with boundary
\cite{Oh:98}. 


\begin{thebibliography}{99}
\bibitem{Sch_picture_rev} A related review article can be found in
  \cite{Jac:95}.
\end{thebibliography}
\end{document}